\documentclass[journal]{IEEEtran}
\usepackage{graphicx}
\usepackage{subfigure}
\usepackage{array}
\usepackage{amsmath}
\usepackage{amssymb}
\usepackage{multirow}
\usepackage{cases}
\usepackage{url}
\usepackage{cite}
\usepackage{fancyhdr}
\usepackage{mathrsfs}
\usepackage[justification=centering]{caption}
\usepackage[colorlinks=true,citecolor=black,linkcolor=black,anchorcolor=black,urlcolor=black,bookmarks=false]{hyperref}
\usepackage{epstopdf}
\ifCLASSINFOpdf
\else
\fi
\begin{document}
%
\title{Tensor Oriented No-Reference Light Field Image Quality Assessment}
%
%
%

\author{Wei Zhou,~\IEEEmembership{Student Member,~IEEE}, Likun Shi, Zhibo Chen,~\IEEEmembership{Senior~Member,~IEEE}, and Jinglin Zhang
\thanks{W. Zhou, L. Shi and Z. Chen are with the CAS Key Laboratory of Technology in Geo-Spatial Information Processing and Application System, University of Science and Technology of China, Hefei 230027, China (e-mail: weichou@mail.ustc.edu.cn; slikun@mail.ustc.edu.cn; chenzhibo@ustc.edu.cn).}
\thanks{J. Zhang is with Key Laboratory of Meteorological Disaster, Ministry of Education, Nanjing University of Information Science and Technology, Nanjing 210044, China (e-mail: jinglin.zhang@nuist.edu.cn).}
\thanks{Corresponding author: Zhibo Chen. Wei Zhou and Likun Shi contributed equally to this paper. This work was supported in part by NSFC under Grant U1908209, Grant 61632001, Grant 41775008, and Grant 61702275, and in part by the National Key Research and Development Program of China under Grant 2018AAA0101400.}}

%
%

\markboth{IEEE Transactions on Image Processing}%
{Shell \MakeLowercase{\textit{et al.}}: Bare Demo of IEEEtran.cls for IEEE Journals}
%



\maketitle

\begin{abstract}
Light field image (LFI) quality assessment is becoming more and more important, which helps to better guide the acquisition, processing and application of immersive media. However, due to the inherent high dimensional characteristics of LFI, the LFI quality assessment turns into a multi-dimensional problem that requires consideration of the quality degradation in both spatial and angular dimensions. Therefore, we propose a novel Tensor oriented No-reference Light Field image Quality evaluator (Tensor-NLFQ) based on tensor theory. Specifically, since the LFI is regarded as a low-rank 4D tensor, the principal components of four oriented sub-aperture view stacks are obtained via Tucker decomposition. Then, the Principal Component Spatial Characteristic (PCSC) is designed to measure the spatial-dimensional quality of LFI considering its global naturalness and local frequency properties. Finally, the Tensor Angular Variation Index (TAVI) is proposed to measure angular consistency quality by analyzing the structural similarity distribution between the first principal component and each view in the view stack. Extensive experimental results on four publicly available LFI quality databases demonstrate that the proposed Tensor-NLFQ model outperforms state-of-the-art 2D, 3D, multi-view, and LFI quality assessment algorithms.
\end{abstract}


%
\IEEEpeerreviewmaketitle

\section{Introduction}
%
%
%
%
\IEEEPARstart{A}{s} an important medium for human visual perception, light enables humans to effectively perceive the spatial, color, form and dynamic changes of our environment. Conventional media modalities such as 2D images mainly consider the intensity information of radiance, which can only provide a two-dimensional sense of presence. Different from traditional image capturing formats, light field content records both radiation intensity and direction information of light rays in the free space, thus providing an enhanced immersive experience. Considering the abundant spatial and angular information of the light field, its processing and application have attracted widespread attention in past decades. However, these operations inevitably introduce heterogeneous artifacts, resulting in the degradation of the perceptual quality for light field content \cite{wu2017light}. Therefore, monitoring the perceptual quality of light field content is critical to better guiding the procedure of light field acquisition, processing and application techniques.

To facilitate the recording and processing of light field content, a 4D function based on the assumption that the light ray radiance is monochromatic, time-invariant and constant along a straight line is adopted to represent light field data \cite{levoy1996light,gortler1996lumigraph}. Specifically, light field is parameterized by four coordinates $\mathbf{L}(s,t,x,y)$, where the $s, t$ dimensions are angular dimensions and $x, y$ dimensions denote spatial dimensions. When a 4D light field image (LFI) is captured by Lytro Illum \cite{lytro_cam}, each view in the LFI is called a sub-aperture image (SAI). Due to the high dimensional characteristics of LFI, its quality is influenced by different dimension of impairments than that of traditional media. Therefore, it is necessary to analyze the specific factors in LFI quality assessment. The existing research works based on subjective evaluation \cite{paudyal2017towards,viola2017comparison,adhikarla2017towards,shi2018light,kara2016effect} found that the LFI quality assessment needs to consider from these three aspects, namely spatio-angular resolution, spatial quality, and angular consistency. Specifically, spatio-angular resolution refers to the number of SAIs in a LFI and the resolution of a SAI. Spatial quality indicates the quality of SAIs and angular consistency measures the visual coherence between SAIs. Since the spatio-angular resolution is usually determined by the acquisition devices, this paper focuses on the effects of spatial quality and angular consistency.



\begin{figure*}[!htb]
	\centering
	\includegraphics[width=18cm]{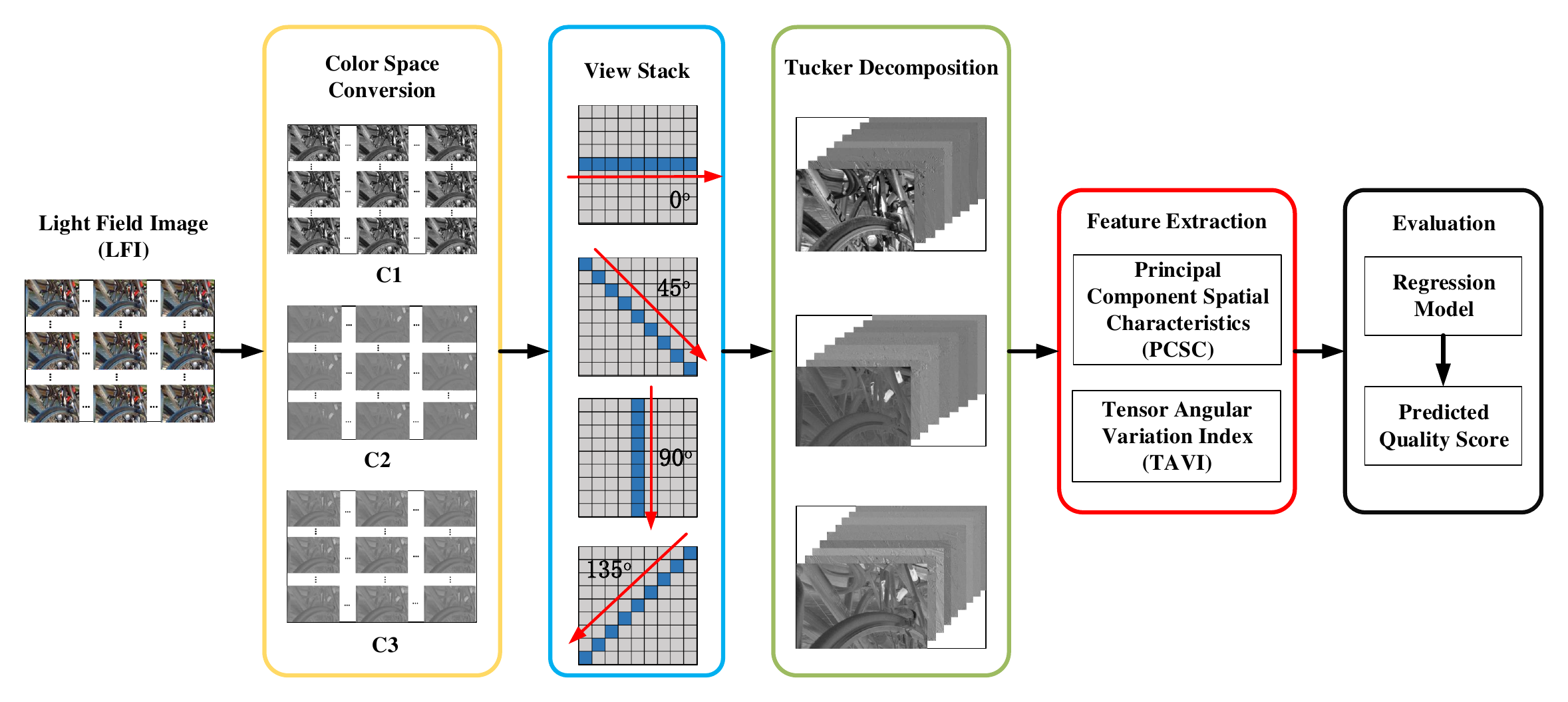}
	\caption{Flow diagram of the proposed Tensor-NLFQ model.}
	\label{figdiagram}
\end{figure*}

Although subjective evaluation is an effective way to understand human behavior and provides reliable image quality scores, it is resource and time consuming without the possibility to be applied for practical applications. Therefore, an effective objective LFI quality assessment model is necessary. Conventionally, image quality assessment (IQA) models can be roughly classified into three categories based on the availability of original reference image information: full-reference (FR), reduced-reference (RR) and no-reference (NR).

However, most of these objective models ignore the intrinsic high dimensional characteristics of LFI. In recent works, the tensor theory has been successfully applied to many fields of computer vision, such as compression and recognition \cite{kolda2009tensor}. Mathematically, a LFI belongs to a 4D tensor. Therefore, the tensor theory can effectively describe the characteristics and distributions in the high-dimensional space. Moreover, these aforementioned methods are designed to extract features in the luminance domain. Although luminance is considered as a dominant factor for understanding the human visual perception \cite{lin2011perceptual}, luminance-based IQA methods may be sub-optimal because they underestimate visual interference caused by color distortion, especially the significant differences in the colors of different SAIs. In addition, existing methods neglect the impact of angular consistency in diverse orientations \cite{8632960,shi2019belif} or only consider horizontal angular consistency \cite{fang2018light,huang2018new}. Since the LFI is an image array, the relationship between each SAI and the adjacent SAI can reflect the LFI angular consistency. Generally, a SAI has eight adjacent SAIs that correspond to angular consistency in four orientation, namely horizontal, vertical, left diagonal, and right diagonal orientations.

In this paper, based on the tensor theory, we proposed a novel Tensor oriented No-reference Light Field image Quality evaluator (Tensor-NLFQ), which considers both luminance and chrominance effects, as well as the impact of angular consistency in diverse directions on the LFI quality. Specifically, the SAIs in RGB are first converted into CIELAB color space, which contains one luminance and two chrominance channels. Second, to comprehensively capture the degradation of LFI angular consistency, view stacks are generated along four orientations. Third, the Tucker decomposition is employed to reduce the angular dimensional of view stacks and obtain the first principal component as the most important dimensionality reduced image. Fourth, considering that the LFI quality is affected by both the spatial quality and angular consistency, we propose the Principal Component Spatial Characteristics (PCSC) for measuring the spatial quality including two key aspects: \romannumeral 1) the naturalness distribution of individual and mutual color channels is extracted to measure the global distortion; \romannumeral 2) local frequency distribution is used to capture local spatial quality degradation. In addition, we propose the Tensor Angular Variation Index (TAVI) to measure the angular consistency, which is computed by analyzing the structural similarity distribution between the first principal component and each view in the view stack. Our experimental results show that the performance of our proposed model correlates well with human visual perception and achieves the state-of-the-art performance. The source codes of Tensor-NLFQ will be available online for public research usage $\footnote{\url{http://staff.ustc.edu.cn/~chenzhibo/resources.html}}$.

The remainder of this paper is organized as follows. Section \uppercase\expandafter{\romannumeral 2} introduces the related work. In Section \uppercase\expandafter{\romannumeral 3}, we present the details of the proposed model. We then illustrate the experimental results in Section \uppercase\expandafter{\romannumeral 4}. Finally, Section \uppercase\expandafter{\romannumeral 5} concludes our paper

\section{Related Work}
The FR IQA approaches utilize the complete reference image information and measure the difference between reference and distorted images. Among a variety of 2D FR IQA methods, structure similarity between reference and distorted images is measured in structural similarity (SSIM) \cite{wang2004image}, and several of its variants have been proposed, i.e. multi-scale SSIM (MS-SSIM) \cite{wang2003multiscale}, feature similarity (FSIM) \cite{zhang2011fsim}, information content weighted SSIM (IW-SSIM) \cite{wang2011information}, and so on \cite{wu2013perceptual,zhang2014vsi,xue2014gradient}. The information fidelity criterion (IFC) \cite{sheikh2005information} and visual information fidelity (VIF) \cite{sheikh2006image} measure the degree of information loss of the distorted images relative to the reference image. Moreover, the noise quality measure (NQM) \cite{damera2000image} and visual signal-to-noise ratio (VSNR) \cite{chandler2007vsnr} consider the sensitivity of the human visual system (HVS) to different visual signals. Chen \textit{et al.} \cite{chen2013full} proposed a 3D FR IQA algorithm that models the influence of binocular rivalry. For multi-view FR IQA, morphological pyramid decomposition and morphological wavelet decomposition are employed in morphological wavelet peak signal-to-noise ratio (MW-PSNR) \cite{sandic2015dibr1} and morphological pyramid PSNR (MP-PSNR) \cite{sandic2015dibr,sandic2016multi}, respectively. The 3D synthesized view image quality metric (3DSwIM) \cite{battisti2015objective} is based on the comparison of statistical features from wavelet subbands.

The RR IQA algorithms utilize partial information of the reference image for quality assessment, which is exploited when the reference information is transmitted at low bandwidth, such as \cite{wang2005reduced,wang2011reduced,rehman2012reduced}. The NR IQA methods measure distorted image quality without needing the original image, which is more applicable in most real-world scenarios. For example, natural scene statistics from different domains are extracted to predict 2D image quality \cite{moorthy2011blind,saad2012blind,mittal2012no,mittal2012making,ghadiyaram2017perceptual}. For 3D NR IQA, binocular vision theory and depth perception are adopted in several methods \cite{chen2013no,liu2017binocular,chen2018blind}. Gu \textit{et al.} \cite{gu2018model} proposed a multi-view NR IQA algorithm named  autoregression (AR)-plus thresholding (APT) that employs the AR-based local image description. However, none of the aforementioned schemes consider the intrinsic high dimensional characteristics of LFI, especially the distortion caused by angular consistency. Therefore, it is important and necessary to design a new light-field-specific metric.

In the literature, several LFI quality assessment models have been proposed. Fang \textit{et al.} \cite{fang2018light} proposed a FR LFI quality assessment method that measures the gradient magnitude similarity of reference and distorted epipolar plane images. Huang \textit{et al.} \cite{huang2018new} also proposed a FR LFI quality assessment algorithm, which is based on dense distortion curve analysis and scene information statistics. The light field image quality assessment metric (LF-IQM) \cite{8632960} is a RR LFI quality assessment metric that assumes the depth map quality is closely related to the LFI overall quality and measures the structural similarity between original and distorted depth maps to predict the perceived LFI quality. However, Fang \cite{fang2018light} and LF-IQM \cite{8632960} ignore the texture information of SAI, which result in the insufficient measurement of the LFI spatial quality. Furthermore, the performance of the LF-IQM is significantly affected by the adopted depth estimation algorithms. Additionally, in most cases, the pristine image is not available, thus NR LFI quality assessment methods are desired. To the best of our knowledge, our previous work \cite{shi2019belif} propose the only NR LFI quality assessment metric called blind quality evaluator of light field image (BELIF), which utilizes binocular vision features for measuring the spatial quality and angular consistency. The differences between BELIF and our proposed Tensor-NLFQ are: 1) The BELIF ignores the effect of chrominance information. However, the proposed Tensor-NLFQ method jointly considers both the influence of luminance and chrominance on perceptual quality; 2) The BELIF neglects the effect of local changes in the angular consistency of each direction, while our proposed Tensor-NLFQ exploits the angular consistency of each direction to evaluate the light field image quality.

\begin{figure}
	\centering
	\begin{minipage}{0.49\linewidth}
		\centerline{\includegraphics[width=4.3cm]{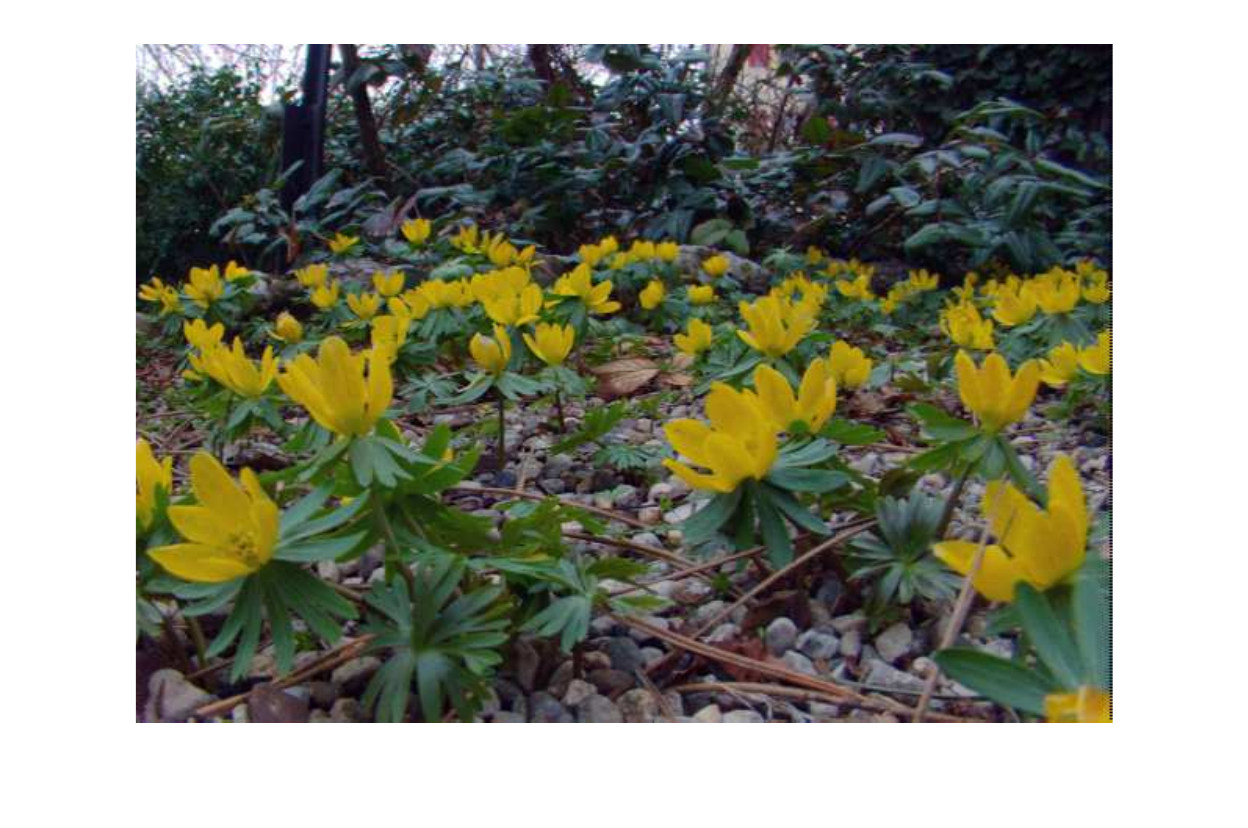}}
		\centerline{(a)}
	\end{minipage}
	\begin{minipage}{0.49\linewidth}
		\centerline{\includegraphics[width=4.3cm]{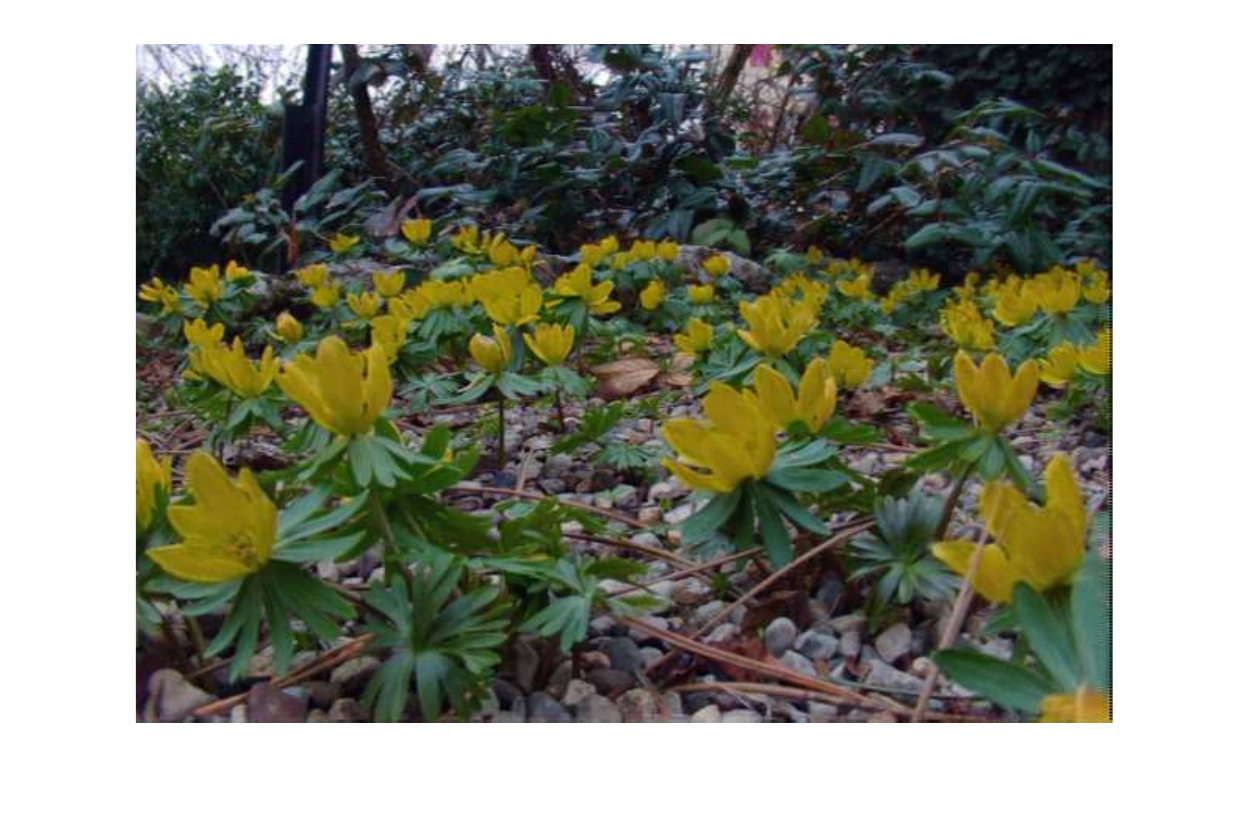}}
		\centerline{(b)}
	\end{minipage}
	\begin{minipage}{0.49\linewidth}
		\centerline{\includegraphics[width=4.3cm]{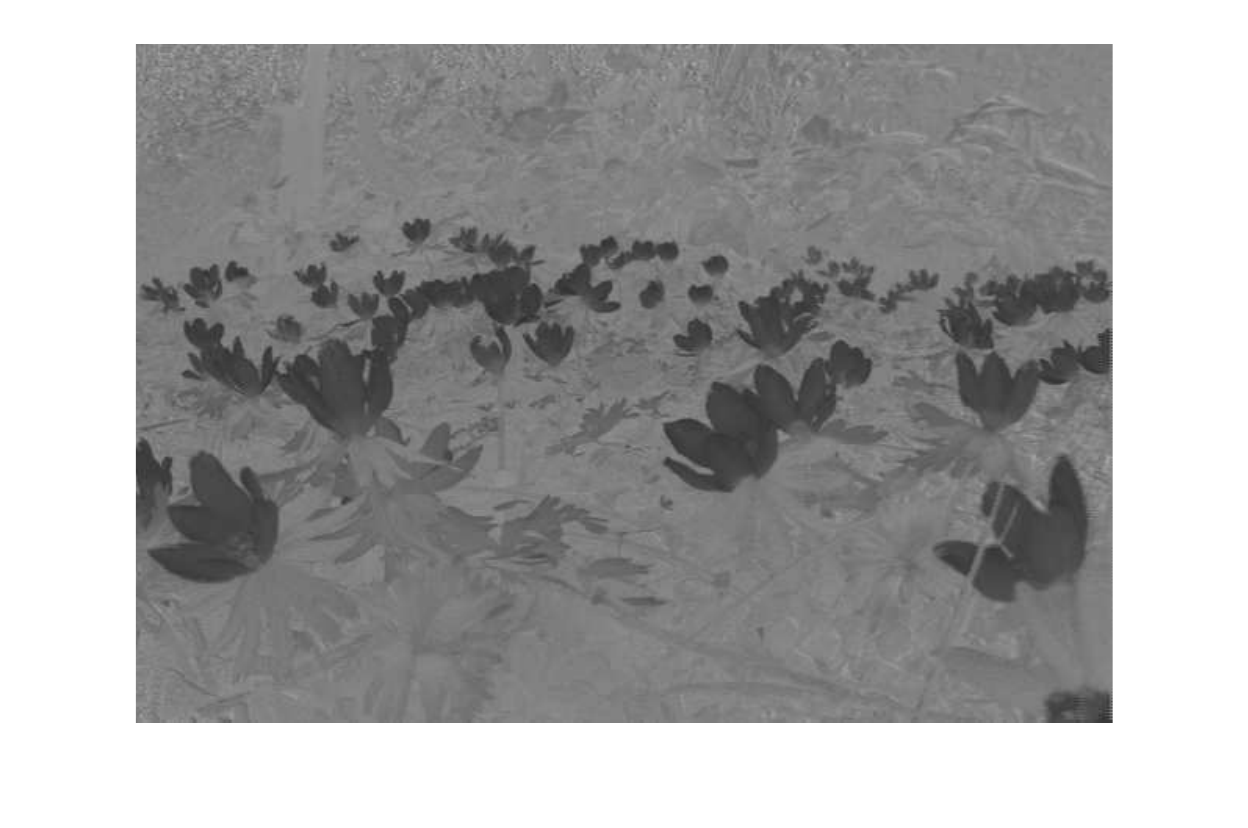}}
		\centerline{(c)}
	\end{minipage}
	\begin{minipage}{0.49\linewidth}
		\centerline{\includegraphics[width=4.3cm]{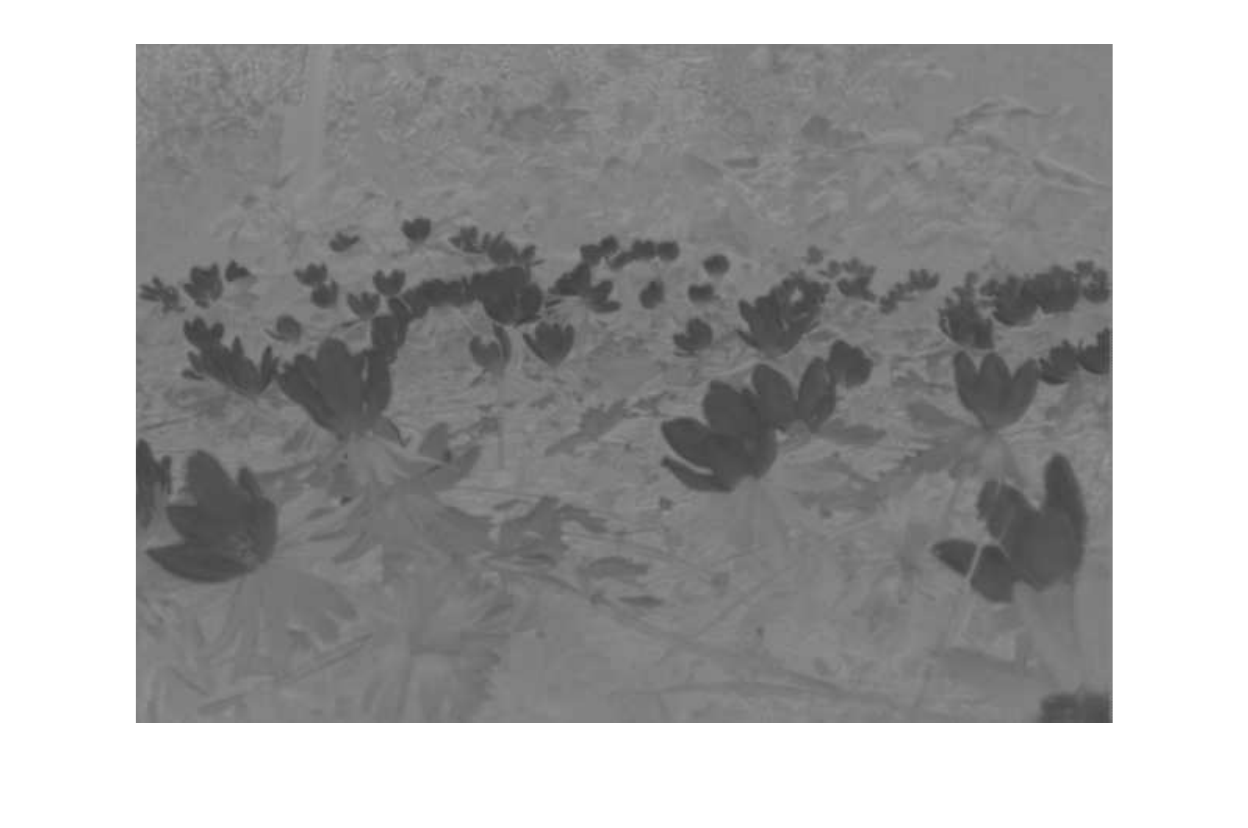}}
		\centerline{(d)}
	\end{minipage}
	\begin{minipage}{0.49\linewidth}
		\centerline{\includegraphics[width=4.3cm]{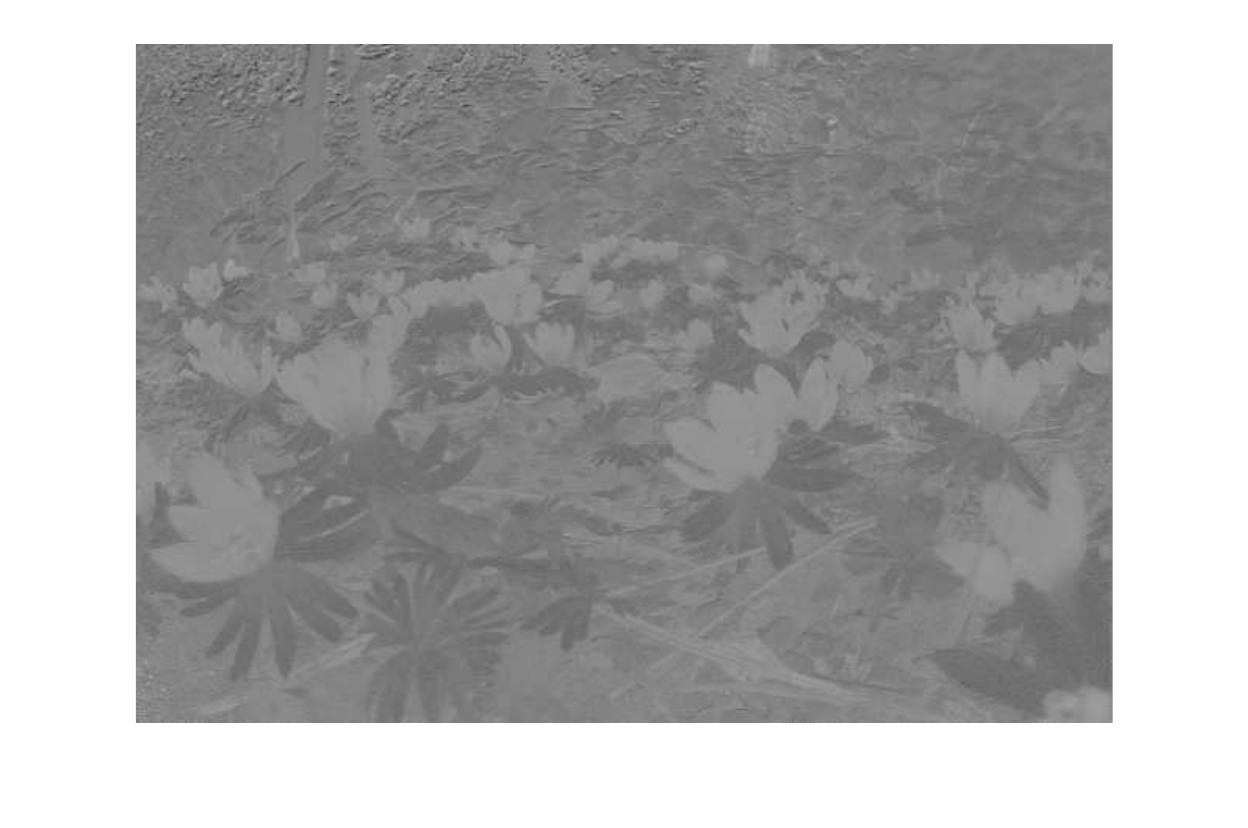}}
		\centerline{(e)}
	\end{minipage}
	\begin{minipage}{0.49\linewidth}
		\centerline{\includegraphics[width=4.3cm]{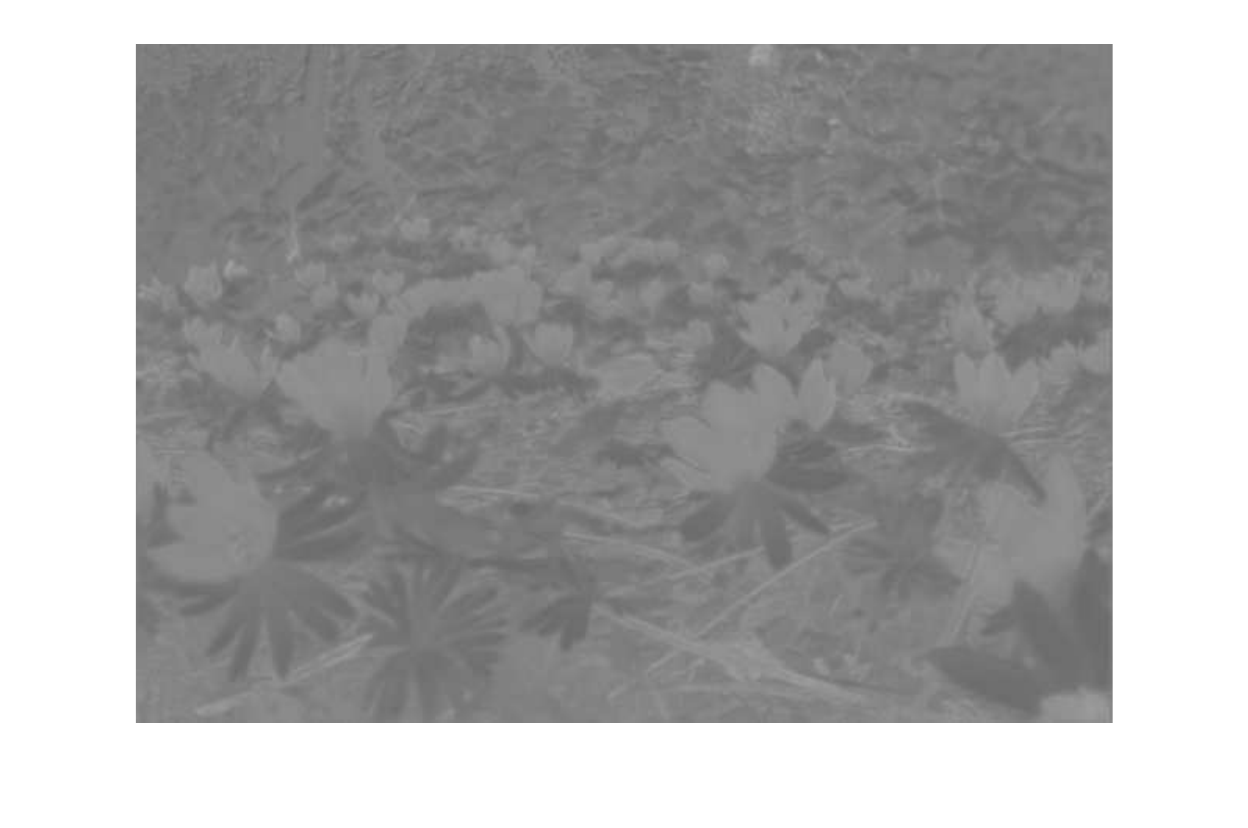}}
		\centerline{(f)}
	\end{minipage}
	\caption{(a-b) Two horizontal adjacent distorted SAIs from Win5-LID database \cite{shi2018light}; (c-d) The corresponding chrominance $a^*$ of (a-b); (e-f) The corresponding chrominance $b^*$ of (a-b).}
	\label{figcolorsai}
\end{figure}

\section{Proposed Method}
The framework of Tensor-NLFQ algorithm is illustrated in Fig. \ref{figdiagram}. First, we convert SAIs in RGB into CIELAB color space. Second, we exploit Tucker decomposition along angular dimension to generate the principal components of view stacks in diverse orientations. Third, the PCSC and TAVI are extracted to measure the degradation of spatial quality and angular consistency, respectively. Finally, we utilize the regression model to predict the perceptual LFI quality.

\begin{figure*}[!htb]
	\centering
	\begin{minipage}{0.25\linewidth}
		\centerline{\includegraphics[width=4cm]{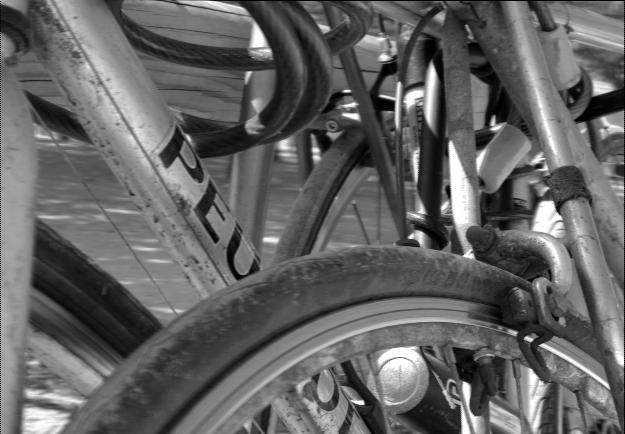}}
		\centerline{}
		\centerline{\includegraphics[width=4cm]{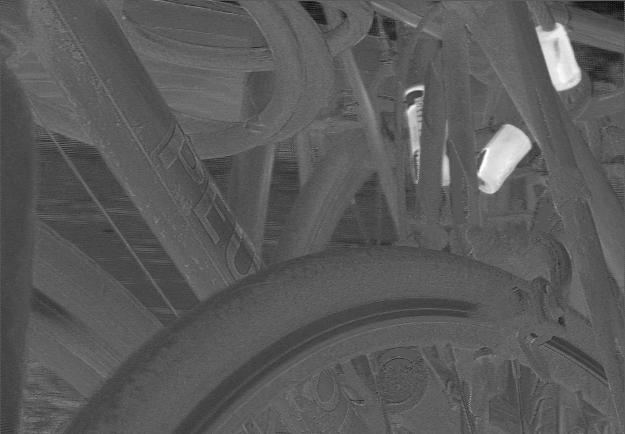}}
		\centerline{}
		\centerline{\includegraphics[width=4cm]{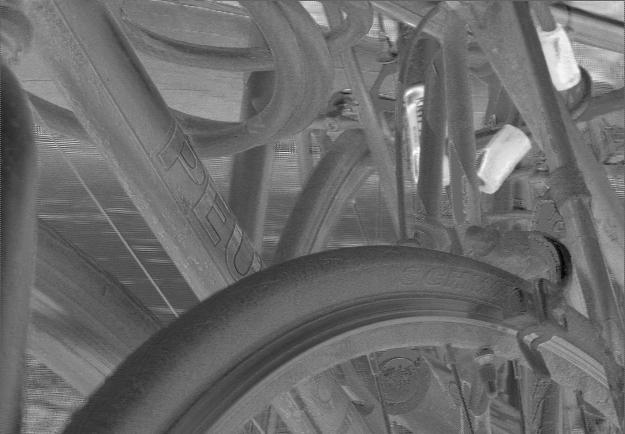}}
		\centerline{(a)}
	\end{minipage}
	\begin{minipage}{0.25\linewidth}
		\centerline{\includegraphics[width=4cm]{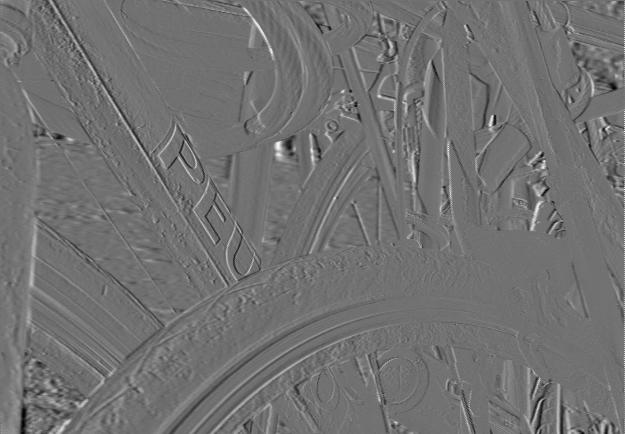}}
		\centerline{}
		\centerline{\includegraphics[width=4cm]{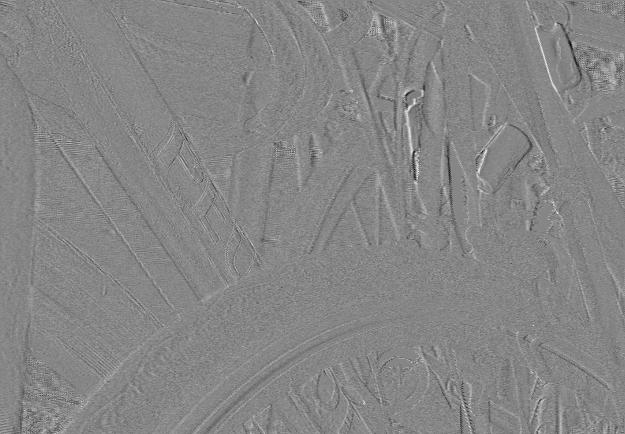}}
		\centerline{}
		\centerline{\includegraphics[width=4cm]{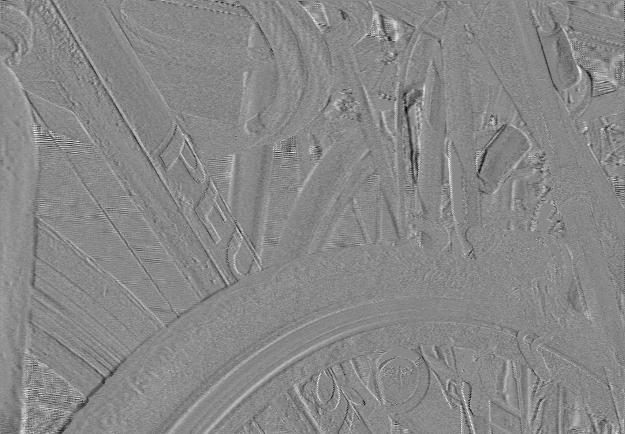}}
		\centerline{(b)}
	\end{minipage}	
	\begin{minipage}{0.25\linewidth}
		\centerline{\includegraphics[width=4cm]{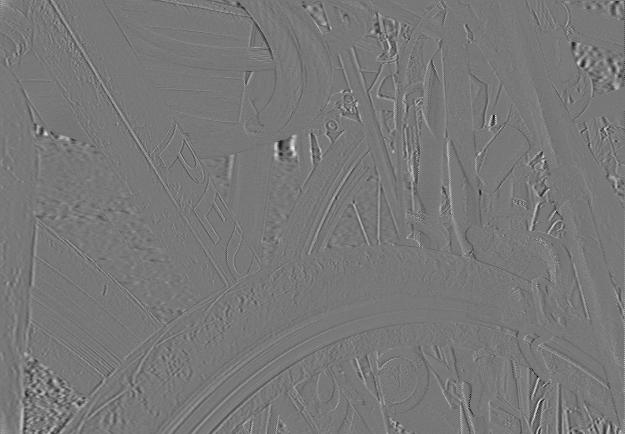}}
		\centerline{}
		\centerline{\includegraphics[width=4cm]{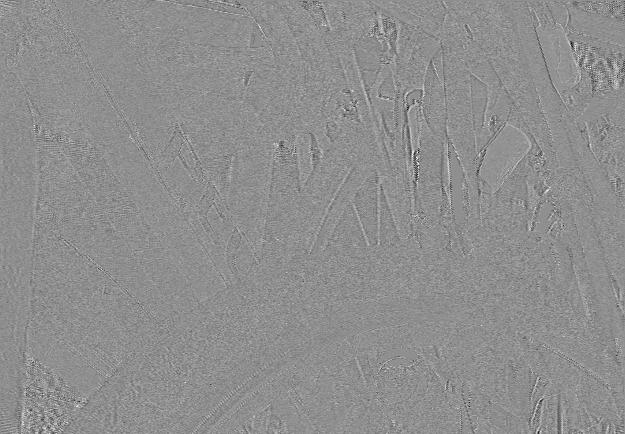}}
		\centerline{}
		\centerline{\includegraphics[width=4cm]{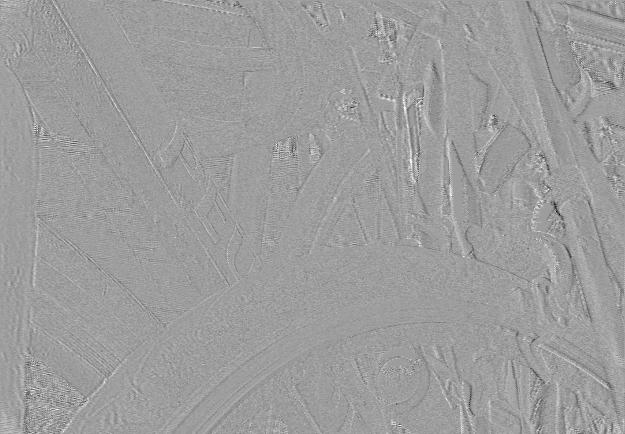}}
		\centerline{(c)}
	\end{minipage}	
	\begin{minipage}{0.22\linewidth}
		\centerline{\includegraphics[width=3.8cm]{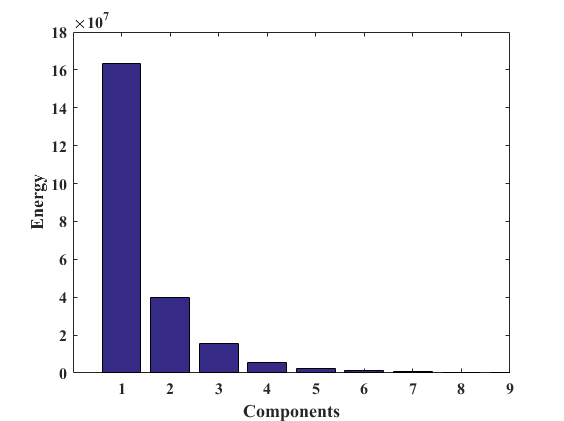}}
		\centerline{}
		\centerline{\includegraphics[width=3.8cm]{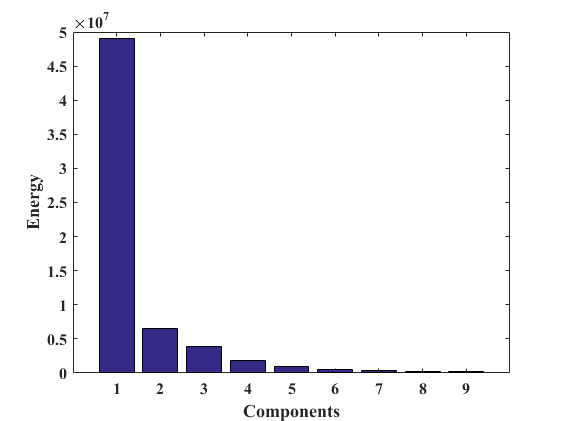}}
		\centerline{}
		\centerline{\includegraphics[width=3.8cm]{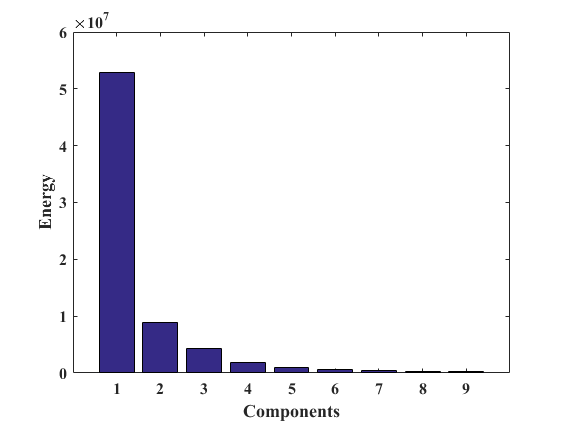}}
		\centerline{(d)}
	\end{minipage}		
	\caption{Tucker decomposition components and their energy histograms. The horizontal view stack is taken as an example, which contains 9 horizontal stacks. And the angular resolution of each stack is 9. The top, middle and bottom rows denote the components of $\mathscr{C}_{1}^{0^\circ}$, $\mathscr{C}_{2}^{0^\circ}$ and $\mathscr{C}_{3}^{0^\circ}$, respectively. (a) the first principal component; (b) the second principal component; (c) the third principal component; (d) energy distribution of the corresponding decomposition components.}
	\label{figTuckerCom}
\end{figure*}

\subsection{Color Space Conversion}
As an important and dense natural visual cue, color information helps the human brain to achieve both low-level and high-level visual perception. Extensive research works have been conducted towards understanding the effects of luminance and chrominance on image quality \cite{lee2015towards,temel2016csv,lee2016toward,ghadiyaram2017perceptual}. These works prove that the chrominance information has a promising gain for image quality evaluation. Therefore, it is reasonable to use the color space information to evaluate LFI spatial quality. Furthermore, in our previous work \cite{shi2018light}, we have found that if there exists a significant difference in the color of different SAIs, this may destroy LFI angular consistency. Fig. \ref{figcolorsai} shows two horizontal adjacent SAIs with reconstruction artifacts selected from Win5-LID database \cite{shi2018light} and the corresponding two chrominance components. We can see that there exist color differences in the SAIs and the two chrominance components are differentiable, which indicate that color information can measure the deterioration of LFI angular consistency.

To better approximate color perception in the HVS, the color SAIs of each LFI are transformed into the perceptually relevant CIELAB color space with one luminance ($L^*$) and two chrominance ($a^*$ and $b^*$) channels optimized for quantifying perceptual color difference and more compatible with human perception \cite{rajashekar2010perceptual}. Specifically, the luminance $L^*$ represents color lightness from black to white. Moreover, $a^*$ indicates the position between red/magenta and green, while $b^*$ represents the position between yellow and blue. Therefore, one luminance map array ($C_1$) and two chrominance map arrays ($C_2$ and $C_3$) can be obtained, as shown in the yellow box in Fig. \ref{figdiagram}. Meanwhile, $C_1$, $C_2$ and $C_3$ have the same spatial resolution and angular resolution as the original LFI.

\subsection{View Stack}
In natural, the distribution of light is continuous. However, for practical usage, the LFI is represented as $\mathbf{L}(s,t,x,y)$, where $(s,t)$ indicates the view index and is an integer. Therefore, except for the corner and boundary SAIs, the remaining SAIs have eight adjacent views. Generally, based on the assumption that the angular resolution of LFI is $S\times T$, each SAI has an angular consistency of four orientations, i.e. $0^\circ$, $45^\circ$, $90^\circ$, and $135^\circ$. As shown in the blue box in Fig. \ref{figdiagram}, these angles represent horizontal ($0^\circ$), left diagonal ($45^\circ$), vertical ($90^\circ$), and right diagonal ($135^\circ$) orientations. We then stack the SAIs along four orientations to generate view stack as follows:

\begin{equation}
\centering
C_{n, s}^{0^\circ} = \{C_{n}(s,1,:,:), C_{n}(s,2,:,:),...,C_{n}(s,T,:,:)\},
\end{equation}

\begin{equation}
\centering
C_{n, t}^{90^\circ} = \{C_{n}(1,t,:,:), C_{n}(2,t,:,:),..., C_{n}(S,t,:,:) \},
\end{equation}

\begin{equation}
\begin{split}
C_{n, s+t-1}^{45^\circ} = \{ C_{n}(s,t,:,:),C_{n}(s+1,t+1,:,:), ...,\\
C_{n}(s+min\{S-s, T-t\},t+min\{S-s, T-t\},:,:)\}  ,
\end{split}
\end{equation}

\begin{equation}
\begin{split}
C_{n, s+t-1}^{135^\circ} = \{C_{n}(s,t,:,:),C_{n}(s+1,t-1,:,:) ,...,\\ C_{n}(s+min\{S-s, T-1\},t-min\{S-s, T-1\},:,:)\},
\end{split}
\end{equation}
where $s=1, 2, ..., S$ and $t=1, 2, ..., T$ represent the angular coordinate. $n = 1, 2, 3$ indicate the luminance and two chrominance channels. For the light field image with an angular resolution of $S \times T$, we extract the view stack in four directions as:
1) Containing $S$ horizontal stacks, and the angular resolution of each stack is $T$; 2) Containing $T$ vertical stacks, and the angular resolution of each stack is $S$; 3) Containing $(S+T-1)$ left diagonal stacks, and the angular resolution of each stack increases from 1 to $min\{S, T\}$; 4) Containing $(S+T-1)$ right diagonal stacks, and the angular resolution of each stack increases from 1 to $min\{S, T\}$. For example, the angular resolution is $9 \times 9$ in Win5-LID database, including 9 horizontal stacks, 9 vertical stacks, 17 left diagonal stacks and 17 right diagonal stacks.

\subsection{Tucker Decomposition}
\begin{figure*}
	\centering
	\begin{minipage}{0.31\linewidth}
		\centerline{\includegraphics[width=6cm]{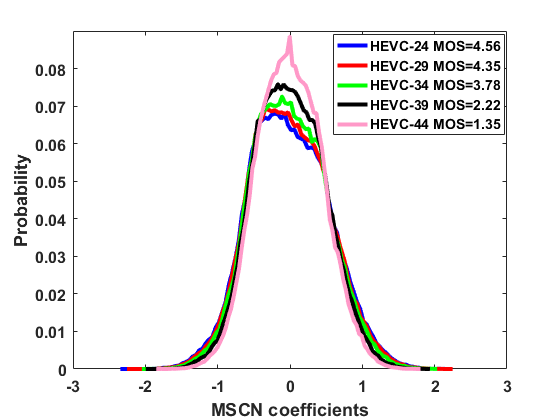}}
		\centerline{(a)}
	\end{minipage}
	\begin{minipage}{0.31\linewidth}
		\centerline{\includegraphics[width=6cm]{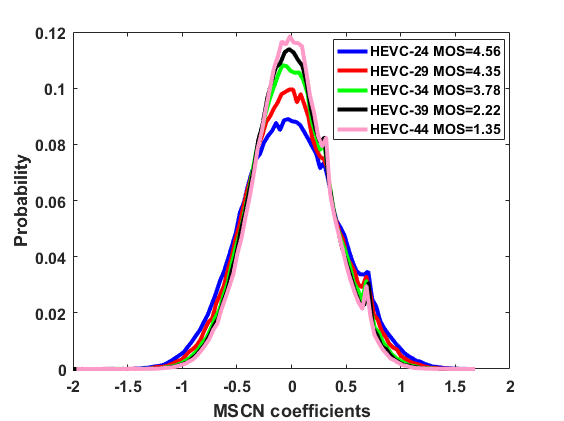}}
		\centerline{(b)}
	\end{minipage}
	\begin{minipage}{0.31\linewidth}
		\centerline{\includegraphics[width=6cm]{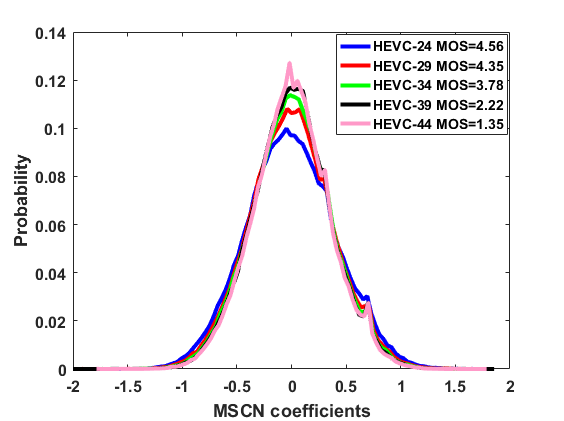}}
		\centerline{(b)}
	\end{minipage}
	\caption{MSCN coefficients for different HEVC compression levels. The HEVC-QP denotes using HEVC standard with specific quantization parameter (QP), where QP=24, 29, 34, 39, and 44. Higher QP represents lower visual quality. (a) luminance MSCN coefficients $\widehat{M_1^{0^\circ}}$; (b) chrominance MSCN coefficients $\widehat{M_2^{0^\circ}}$; (c) chrominance MSCN coefficients $\widehat{M_3^{0^\circ}}$;.}
	\label{figmscn}
\end{figure*}

The view stack is a 3D signal that includes two spatial coordinates and one angular coordinate. We discover that there exists a high texture similarity between different images of the view stack, indicating that there exists a large redundancy in the angular dimension. To alleviate this problem, we first adopt tensor decomposition to remove redundant information from the angular dimension. It should be noted that there exist significant differences between Tucker decomposition and principal component analysis (PCA). According to \cite{kolda2009tensor,lee2014incremental}, the Tucker decomposition can be taken as the higher-order generalizations of PCA or singular value decomposition (SVD). Moreover, the PCA operates on two-dimensional matrices, which vectorizes the image and destroys the spatial structure information of the image. However, the Tucker decomposition decomposes tensors in high-dimensional space that can retain the spatial structure information of the image. Therefore, the Tucker decomposition is used to achieve dimensionality reduction \cite{kolda2009tensor}. It decomposes a tensor into a core tensor multiplied by a matrix along each dimension. In other words, we decompose the three-dimensional light field signal into the core tensor and the principal components of spatial and angular dimensions. For horizontal view stack $C_{n}^{0^\circ}$, we thus have:

\begin{equation}
C_{n}^{0^\circ} \approx \mathscr{G} \times_1 U_{1} \times_2 U_{2} \times_3 U_{3},
\end{equation}
where $\mathscr{G} \in \mathbb{R}^{R_1 \times R_2 \times R_3}$ is the core tensor whose entries illustrate the level of interaction between different components. $U_{1} \in \mathbb{R}^{K_1 \times R_1}$ and $U_{2} \in \mathbb{R}^{K_2 \times R_2}$ are the factor matrices in the spatial dimension. $U_{3} \in \mathbb{R}^{K_3 \times R_3}$ is the angular dimension factor matrix. These matrices are usually orthogonal. In our model, we set $K_n=R_n$, where $n = 1, 2, 3$. The core tensor is trained from each distorted SAI.

Then, for $C_{n}^{0^\circ}$, the angular decomposition components can be obtained by multiplying the core tensor with the factor matrices $U_{1}$ and $U_{2}$ along each mode in the spatial dimension, which can be given by:

\begin{equation}
\mathscr{C}_{n}^{0^\circ} = \mathscr{G} \times_1 U_{1} \times_2 U_{2},
\end{equation}
where $C_{n}^{0^\circ} \approx \mathscr{C}_{n}^{0^\circ} \times_3 U_{3}$. That is, we apply the mode product to the core tensor and the principal components of the spatial dimension, which can obtain angular decomposition components. The purpose of removing $U_{3}$ is to realize the reconstruction of spatial information and obtain the decomposition components of angular dimension. Similar to the computation process of $\mathscr{C}_{n}^{0^\circ}$, we obtain the angular decomposition components $\mathscr{C}_{n}^{45^\circ}$, $\mathscr{C}_{n}^{90^\circ}$ and $\mathscr{C}_{n}^{135^\circ}$ of view stacks in other orientations. Specifically, we utilize the alternating least squares method provided by the tensor toolbox \cite{TTB_Software} to implement the Tucker decomposition. Note that the tensor n-mode product represents different modes of tensor multiplication, which is essentially the multiplication of tensors with different dimensions \cite{kolda2009tensor}.

\begin{figure*}[!htb]
	\centering
	\begin{minipage}{0.23\linewidth}
		\centerline{\includegraphics[width=4.7cm]{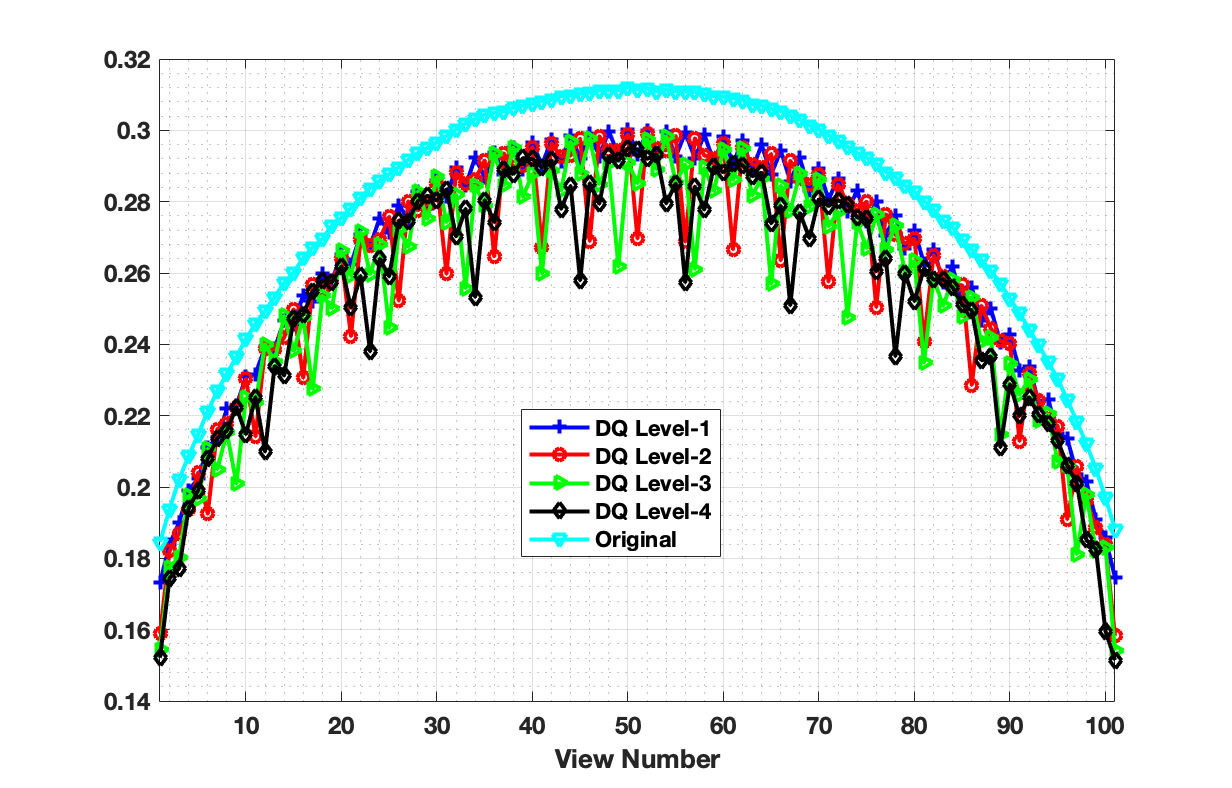}}
		\centerline{}
		\centerline{\includegraphics[width=4.7cm]{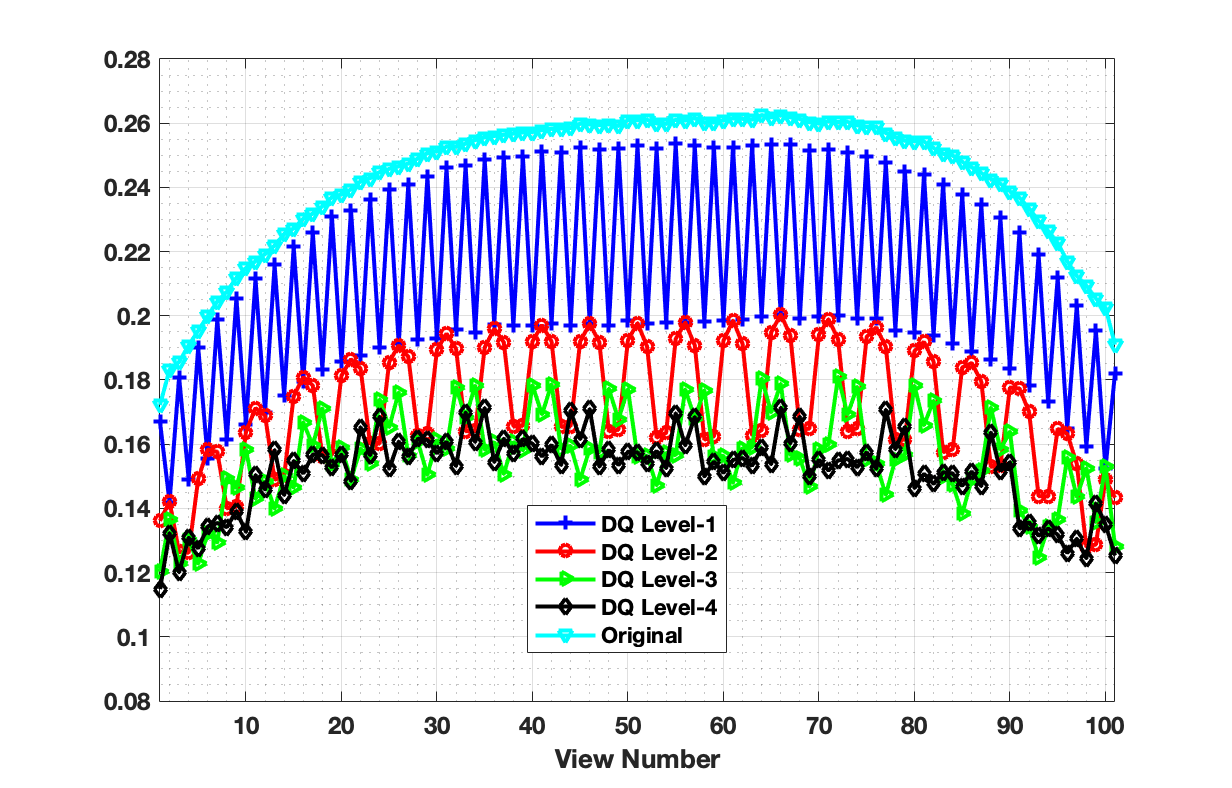}}
		\centerline{}
		\centerline{\includegraphics[width=4.7cm]{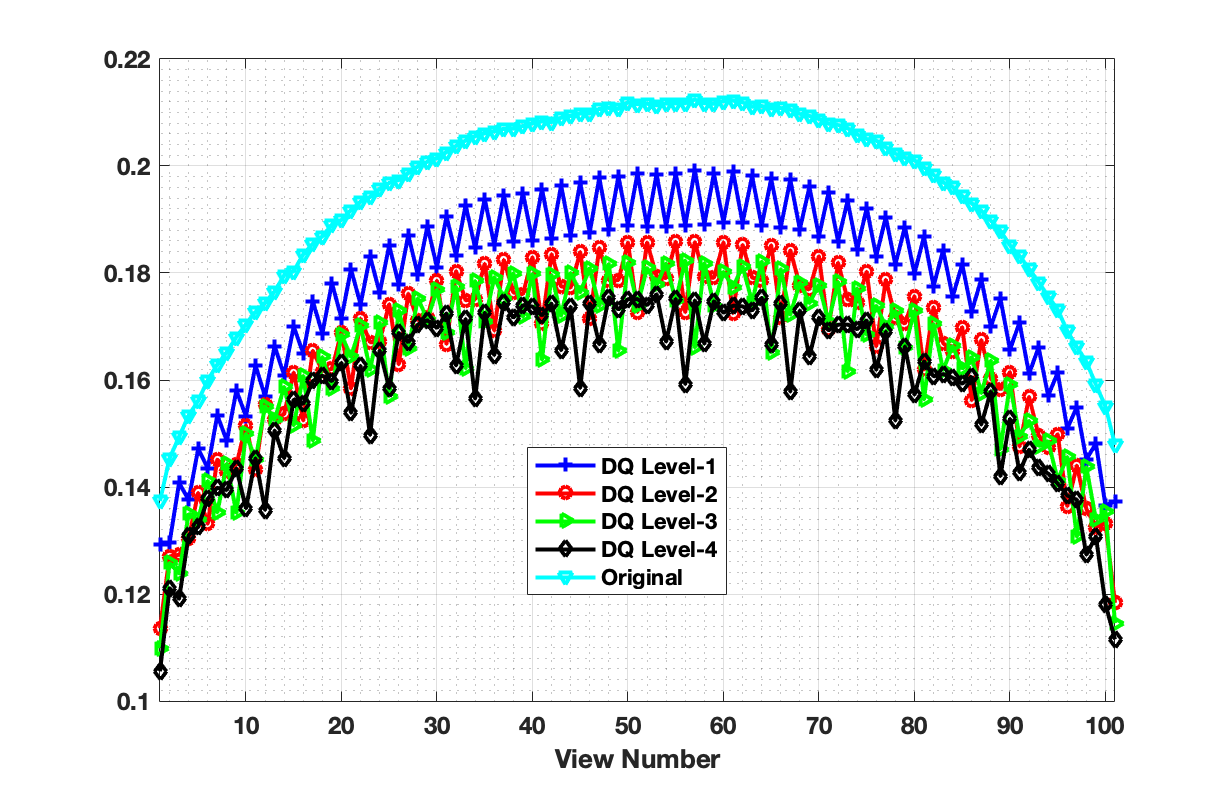}}
		\centerline{(a)}
	\end{minipage}
	\begin{minipage}{0.23\linewidth}
		\centerline{\includegraphics[width=4.7cm]{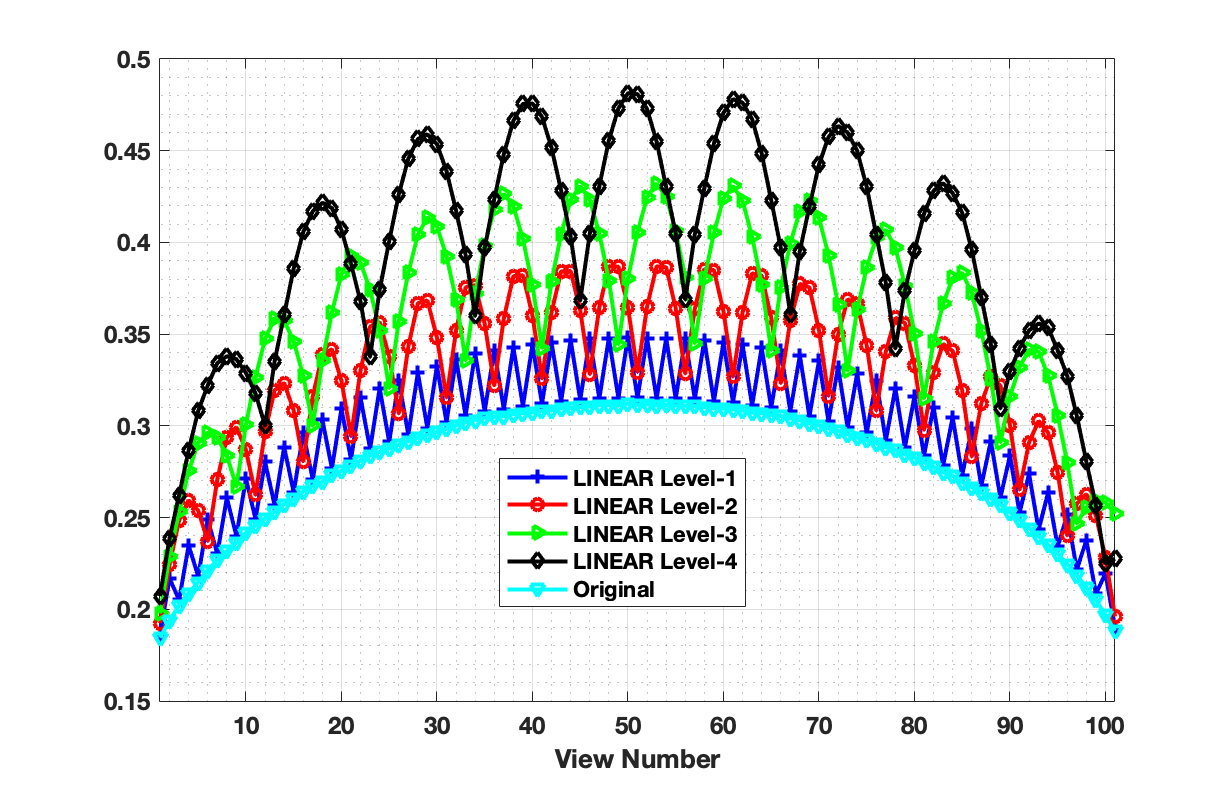}}
		\centerline{}
		\centerline{\includegraphics[width=4.7cm]{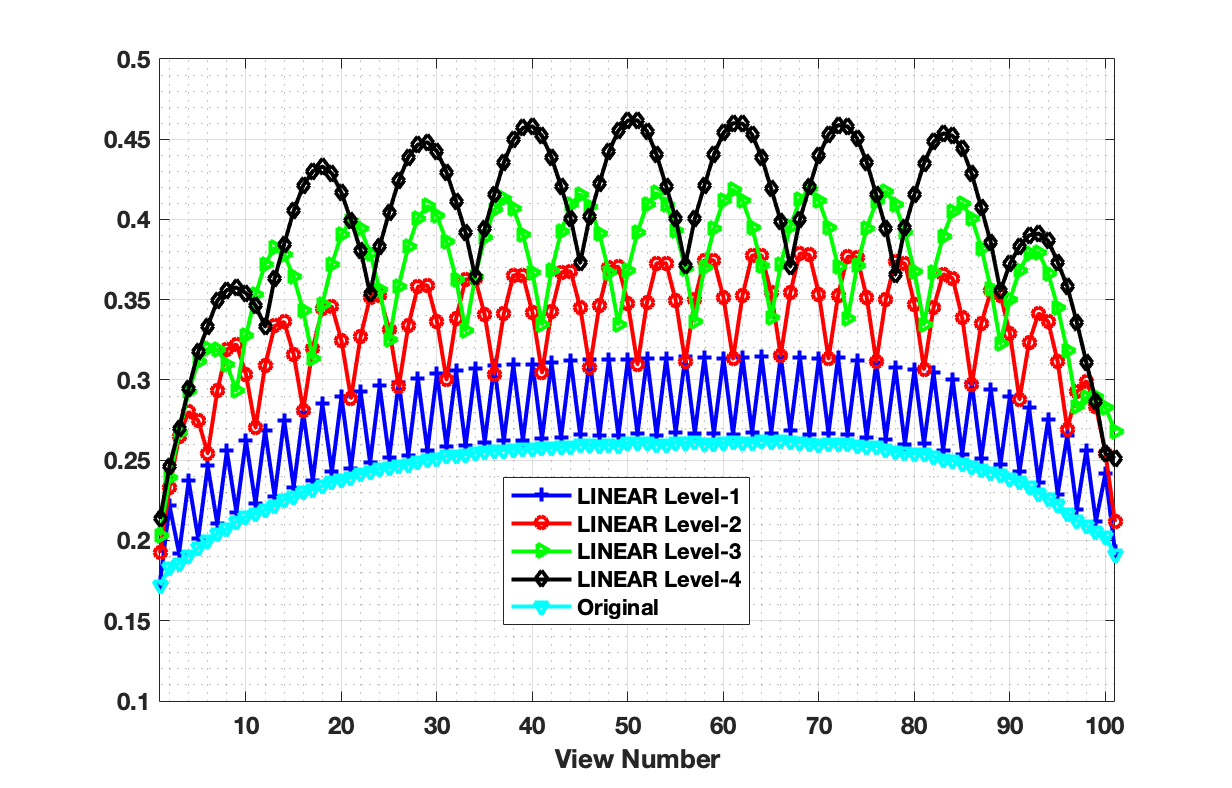}}
		\centerline{}
		\centerline{\includegraphics[width=4.7cm]{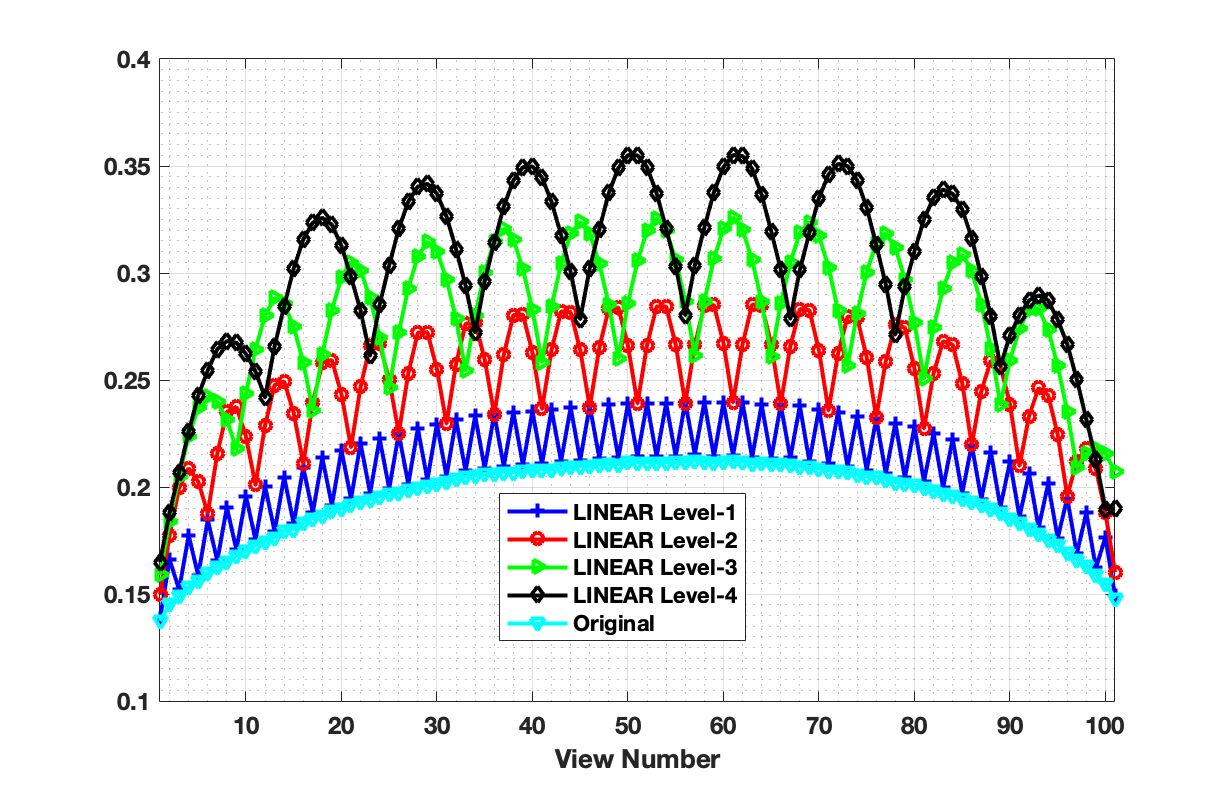}}
		\centerline{(b)}
	\end{minipage}	
	\begin{minipage}{0.23\linewidth}
		\centerline{\includegraphics[width=4.7cm]{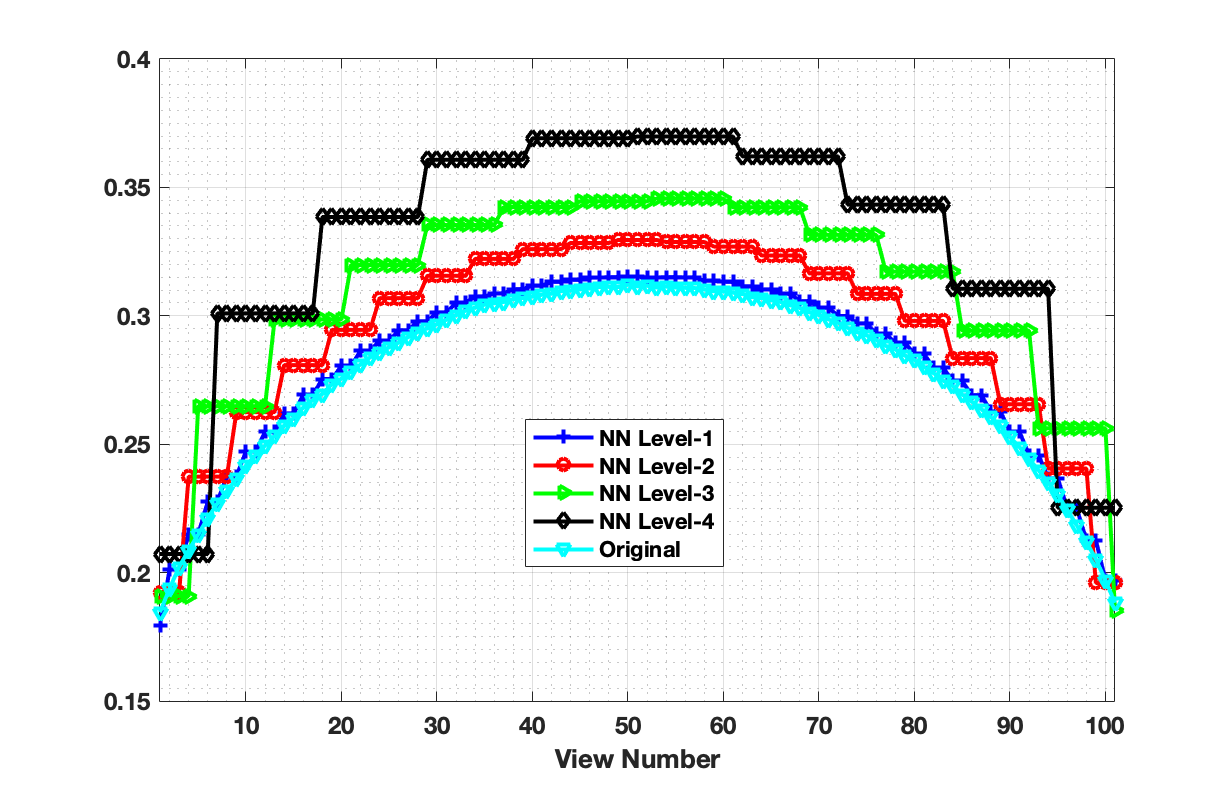}}
		\centerline{}
		\centerline{\includegraphics[width=4.7cm]{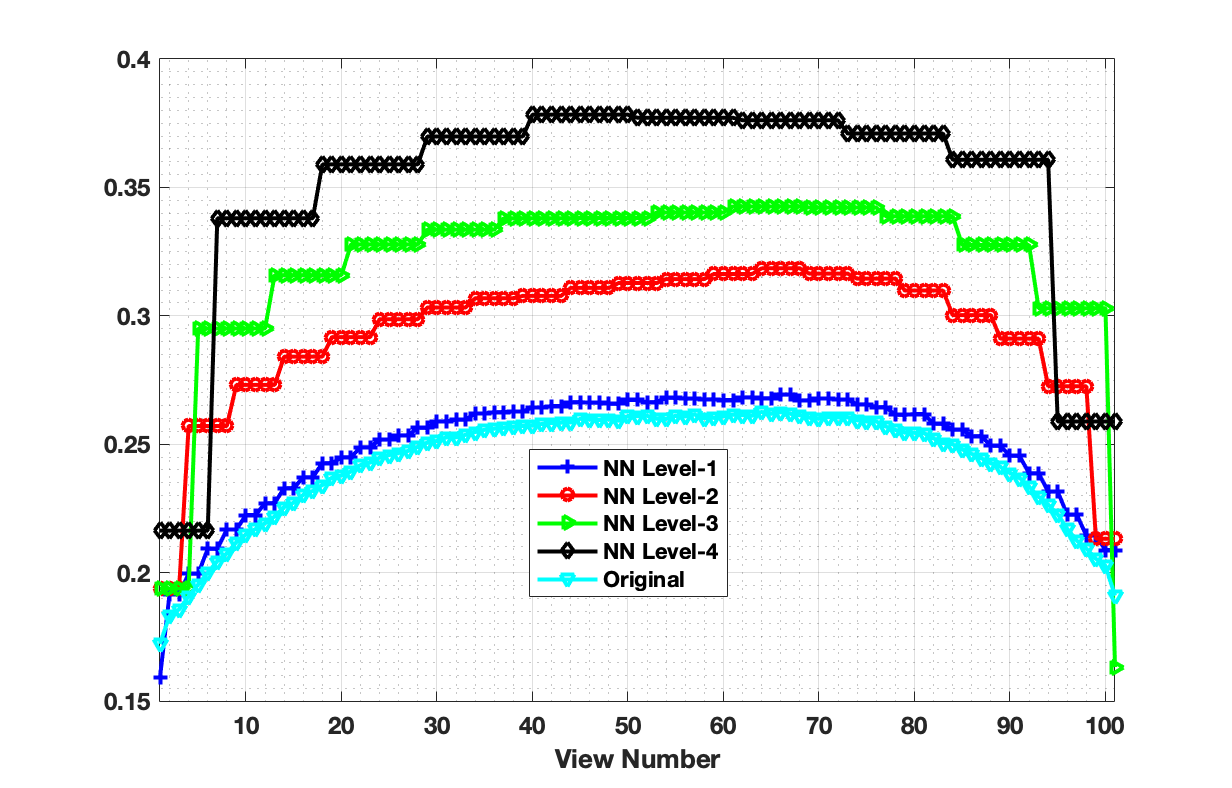}}
		\centerline{}
		\centerline{\includegraphics[width=4.7cm]{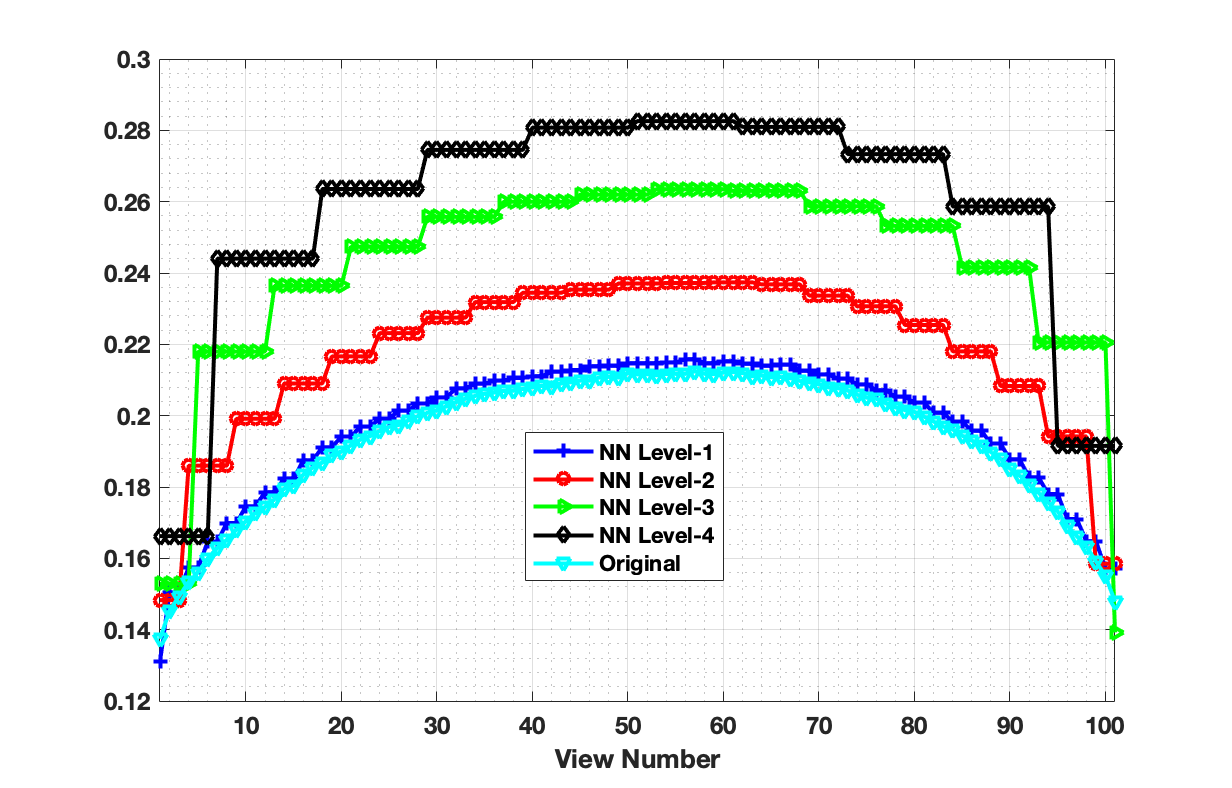}}
		\centerline{(c)}
	\end{minipage}	
	\begin{minipage}{0.23\linewidth}
		\centerline{\includegraphics[width=4.7cm]{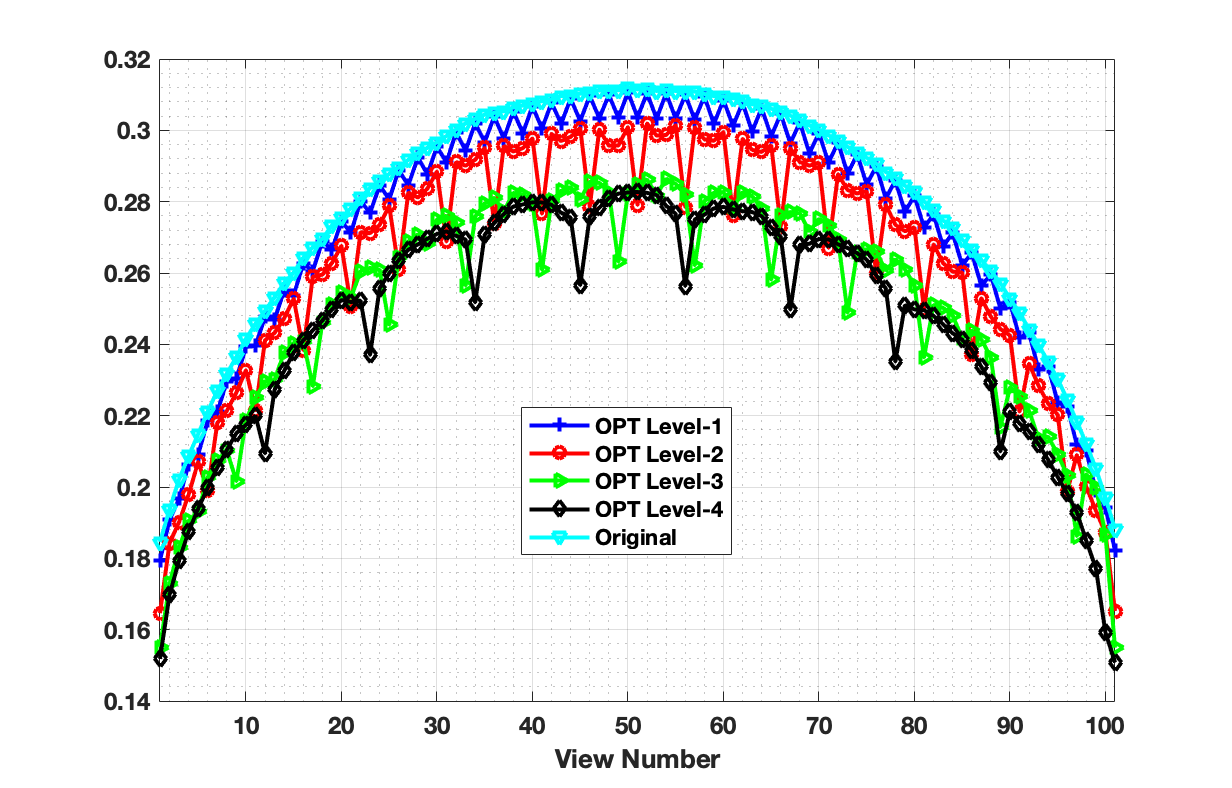}}
		\centerline{}
		\centerline{\includegraphics[width=4.7cm]{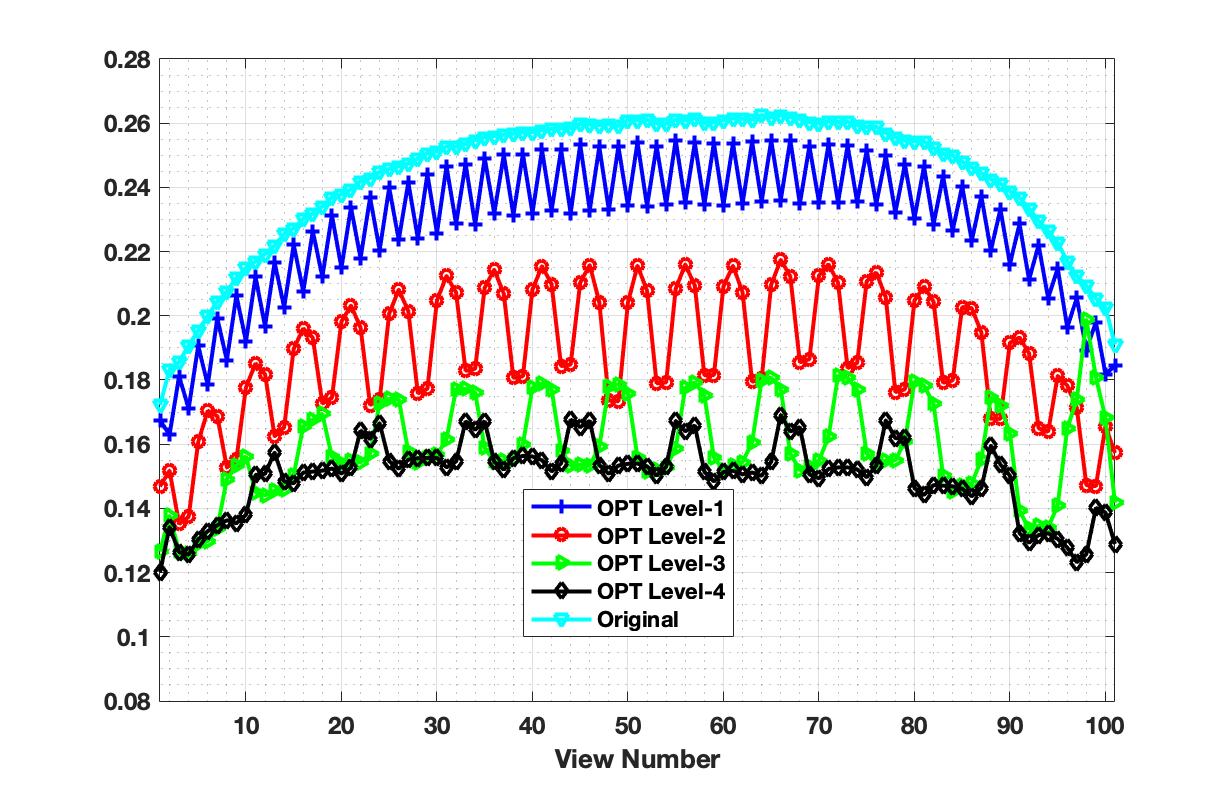}}
		\centerline{}
		\centerline{\includegraphics[width=4.7cm]{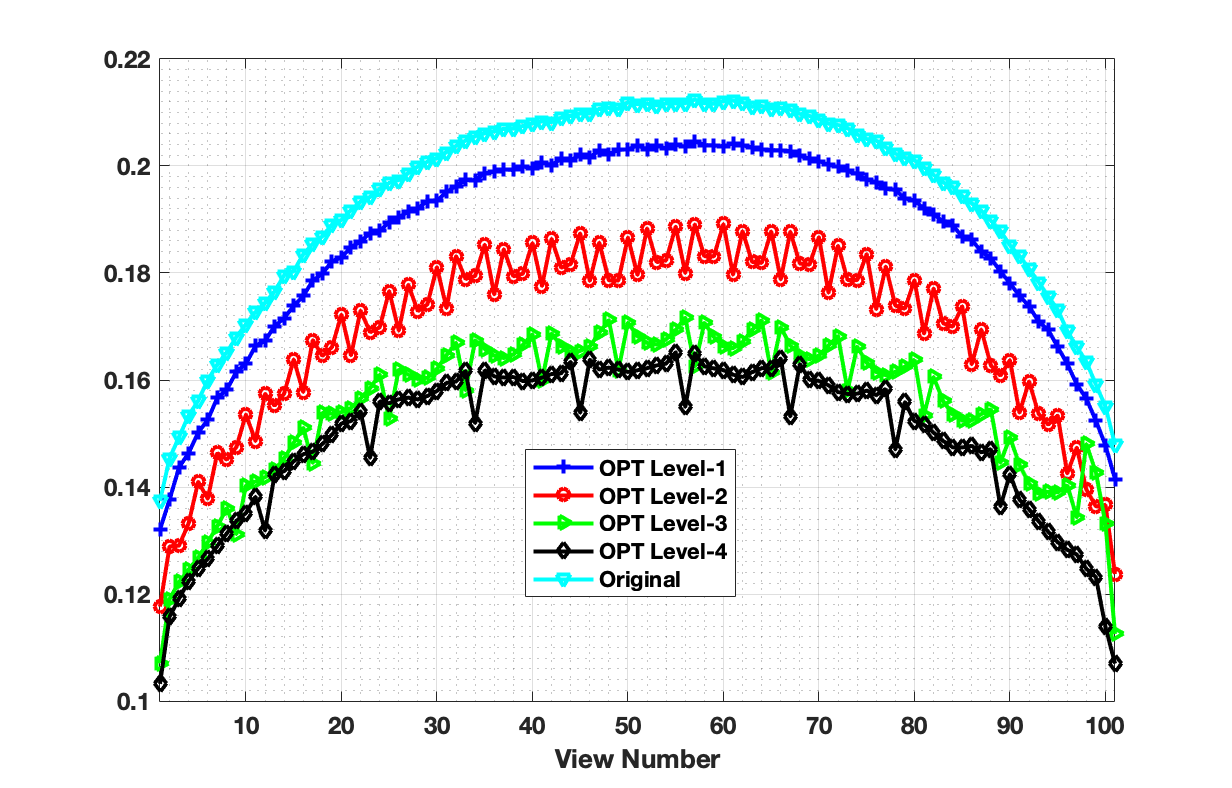}}
		\centerline{(d)}
	\end{minipage}		
	\caption{Structure similarity distribution of horizontal view stack. The top row represents the $\textbf{ss}_{1}^{0^\circ}$ of luminance $L$ channel, while the middle and bottom rows indicate the $\textbf{ss}_{2}^{0^\circ}$ and $\textbf{ss}_{3}^{0^\circ}$ of chrominance $a$ and $b$ channels. (a) Different DQ distortion levels; (b) Different LINEAR distortion levels; (c) Different NN distortion levels; (d) Different OPT distortion levels.}
	\label{figSS}
\end{figure*}

In addition, here the factor matrices in Tucker decomposition, which are defined as the principal components \cite{kolda2009tensor}, represent the stacks in the angular dimension for the decomposed three-dimensional tensor, where the first principal component is the highest energy component and contains fundamental texture information. We select a sample of LFI from Win5-LID database to show the energy distribution of principal components in Fig. \ref{figTuckerCom}, where (a)-(c) illustrates the first three principal components of $\mathscr{C}_{1}^{0^\circ}$, $\mathscr{C}_{2}^{0^\circ}$ and $\mathscr{C}_{3}^{0^\circ}$. Fig. \ref{figTuckerCom} (d) shows the energy histogram distribution of the corresponding decomposition components. Here, the top, middle and bottom rows denote the components of $\mathscr{C}_{1}^{0^\circ}$, $\mathscr{C}_{2}^{0^\circ}$ and $\mathscr{C}_{3}^{0^\circ}$, respectively. Obviously, the texture information and energy mainly concentrate on the first principal component, which represents the basic texture information of the view stack. By quantitative calculation, we find the first principal component contains more than $70\%$ energy, we thus treat it as the most important dimensionality reduced image. We define the first principal component of $\mathscr{C}_{n}^{d}$ as $M_n^{d}$, where $n = 1, 2, 3$ and $d = {0^{\circ}, 45^{\circ}, 90^{\circ}, 135^{\circ}}$.

\subsection{Feature Extraction and Quality Regression}
Since the first principal component contains the basic information about each view stack, it is reasonable to extract features from the first principal component to measure the degradation of LFI spatial quality. Specifically, we first extract the PCSC from the first principal component that utilizes global naturalness and local frequency distribution characteristics to evaluate the distortion in spatial quality. In addition to spatial quality, angular consistency also affects LFI quality. Then, the TAVI is proposed to capture angular consistency distortion by computing the structural similarity between the first principal component and each view in the view stack.


\subsubsection{Principal Component Spatial Characteristic (PCSC)}
In general, the naturalness of an image can be effectively measured by modeling the locally mean subtracted and contrast normalized (MSCN) coefficients \cite{moorthy2011blind,saad2012blind,mittal2012no}. The MSCN has been successfully employed for image processing tasks and can be used to model the contrast-gain masking process in early human vision \cite{mittal2012no,carandini1997linearity}. In our model, MSCN coefficients can be calculated by:

\begin{equation}
\widehat{I}(x,y)=\frac{I(x,y)-\mu(x,y)}{\sigma(x,y)+1},
\end{equation}
where $\widehat{I}(x,y)$ and $I(x,y)$ are the MSCN coefficients and input image (i.e. $M_n^{d}$) values at the spatial position $(x,y)$. $\mu(x,y)$ and $\sigma(x,y)$ stand for the local mean and standard deviation in a local patch centered at $(x,y)$. They are computed as:

\begin{equation}
\mu(x,y)= \sum\limits_{k=-K}^{K} \sum\limits_{l=-L}^{L} z_{k,l} I_{k,l}(x,y)
\end{equation}

\begin{equation}
\sigma(x,y) = \sqrt{\sum\limits_{k=-K}^{K}\sum\limits_{l=-L}^{L} z_{k,l} (I_{k,l}(x,y)-\mu(x,y))^2},
\end{equation}
where $z=\{z_{k,l}|k=-K,...,K,l=-L,...,L\}$ denotes a 2D circularly-symmetric Gaussian weighting function with sampled out 3 standard deviations and rescaled to unit volume. Inspired by \cite{mittal2012no}, we set $K=L=3$ in our implementation.

To measure the LFI spatial quality, we first consider the naturalness distribution of the principal components of luminance and chrominance (i.e. $\widehat{M_1^{d}}$, $\widehat{M_2^{d}}$ and $\widehat{M_3^{d}}$). Fig. \ref{figmscn} presents the distribution of MSCN coefficients for luminance and chrominance principal components with several high efficiency video coding (HEVC) compression levels. The HEVC is the video coding standard promoted by the Joint Collaborative Team on Video Coding in 2013, and it is also named H.265 \cite{sullivan2012overview}. The results show that the distribution of MSCN coefficients are very indicative when the LFI suffers from artifacts. Here, the sample of LFI is selected from the Win5-LID database \cite{shi2018light}. Since the distribution of MSCN coefficients approximates Gaussian distribution and the asymmetric generalized Gaussian distribution (AGGD) further generalizes the GGD \cite{mittal2012no,sharifi1995estimation}, we then utilize the zero-mean AGGD model to qualify the distribution of MSCN coefficients, which can fit the distribution by:

\begin{small}
	\begin{equation}
	f(\chi;\alpha,\sigma_l^2,\sigma_r^2)=\left\{
	\begin{aligned}
	\frac{\alpha}{(\beta_l+\beta_r)\Gamma(\frac{1}{\alpha})}exp(-(\frac{-x}{\beta_l})^\alpha) & & \chi<0 \\
	\frac{\alpha}{(\beta_l+\beta_r)\Gamma(\frac{1}{\alpha})}exp(-(\frac{-x}{\beta_r})^\alpha) & & \chi \geqslant 0,
	\end{aligned}
	\right.
	\end{equation}
\end{small}
where
\begin{equation}
\beta_l=\sigma_l\sqrt{\frac{\Gamma(\frac{1}{\alpha})}{\Gamma(\frac{3}{\alpha})}}  \quad and \quad \beta_r=\sigma_r\sqrt{\frac{\Gamma(\frac{1}{\alpha})}{\Gamma(\frac{3}{\alpha})}},
\end{equation}
and $\alpha$ is the shape parameter controlling the shape of the statistic distribution, while $\sigma_l$ and $\sigma_r$ are the scale parameters of left and right sides, respectively. Moreover, we compute $\eta$ as another feature by:

\begin{equation}
\eta=(\beta_r-\beta_l)\frac{\Gamma(\frac{2}{\alpha})}{\Gamma(\frac{1}{\alpha})}
\end{equation}

In addition, human visual perception is also affected by the combination of luminance and chrominance channels. Therefore, the joint statistics of MSCN coefficients for the principal components under different channels can also be used to measure the deterioration of image quality. Therefore, we utilize multivariate generalized Gaussian distribution (MGGD) \cite{su2014bivariate,sinno2018towards} to fit the joint distribution, which is defined as:

\begin{equation}
\centering
f(\mathbf{x} | \mathbf{M}, \gamma,\varphi)=\frac{1}{|\mathbf{M}|^{\frac{1}{2}}} g_{\gamma,\varphi}\left(\mathbf{x}^{T} \mathbf{M}^{-1} \mathbf{x}\right),
\end{equation}
where $\mathbf{x} \in \mathbb{R}^{N}$ and M is an $N\times N$ symmetric scatter matrix. $\gamma$ and $\varphi$ indicate the scale and shape parameters, respectively. $g_{\gamma,\varphi}()$ is the density generator:

\begin{equation}
\centering
g_{\gamma, \varphi}(\chi)=\frac{\varphi \Gamma\left(\frac{N}{2}\right)}{\left(2^{\frac{1}{\varphi}} \pi \gamma\right)^{\frac{N}{2}} \Gamma\left(\frac{N}{2 \varphi}\right)} e^{-\frac{1}{2}\left(\frac{\chi}{\gamma}\right)^{\varphi}},
\end{equation}
where $\chi \in \mathbb{R}^{+}$ and $\Gamma$ is the digamma function. We adopt the method proposed by Pascal \textit{et al.} \cite{pascal2013parameter} to estimate the parameters of the MGGD model.


Considering that the degradation of LFI spatial quality induces the change in the local distribution of principal components. We extract the local features of principal components for each color channel. Inspired by \cite{saad2012blind}, block-based discrete cosine transform (DCT) is utilized to measure the distribution of local information. Specifically, we adopt the entropy of DCT coefficients without DC value as:

\begin{equation}
\centering
E = -\sum_{l}^{L} \sum_{h}^{H} (p_{lh}log(p_{lh})),
\label{dctentropy}
\end{equation}
where $L$ and $H$ represent the width and height of DCT block, respectively. $p_{lh}$ is the DCT coefficient located in $(l, h)$. Note that we compute the entropy from three aspects, namely the whole DCT block, three frequency bands, and three orientations of the DCT block as \cite{saad2012blind}. Therefore, $\textbf{f}_{PCSC}$ is obtained by concatenating the fitting parameters of AGGD and MGGD as well as three averaged entropy features. Furthermore, the feature dimension of PCSC is 57, where the entropy feature contains 15 dimensions and the feature dimensions for the MSCN based AGGD parameters as well as MGGD parameters are 36 and 6, respectively. There exists significant differences between traditional spatial features such as BRISQUE \cite{mittal2012no} and our proposed $\textbf{f}_{PCSC}$. Specifically, the BRISQUE \cite{mittal2012no} only considers the information distribution of a single luminance space. In addition to the luminance information, we also calculate the distribution of chrominance space. Considering the interaction between luminance and chroma information, we calculate the joint distribution MGGD.

\subsubsection{Tensor Angular Variation Index (TAVI)}

\begin{figure}[!htb]
\begin{minipage}{0.55\linewidth}
  \centerline{\includegraphics[width=4.5cm]{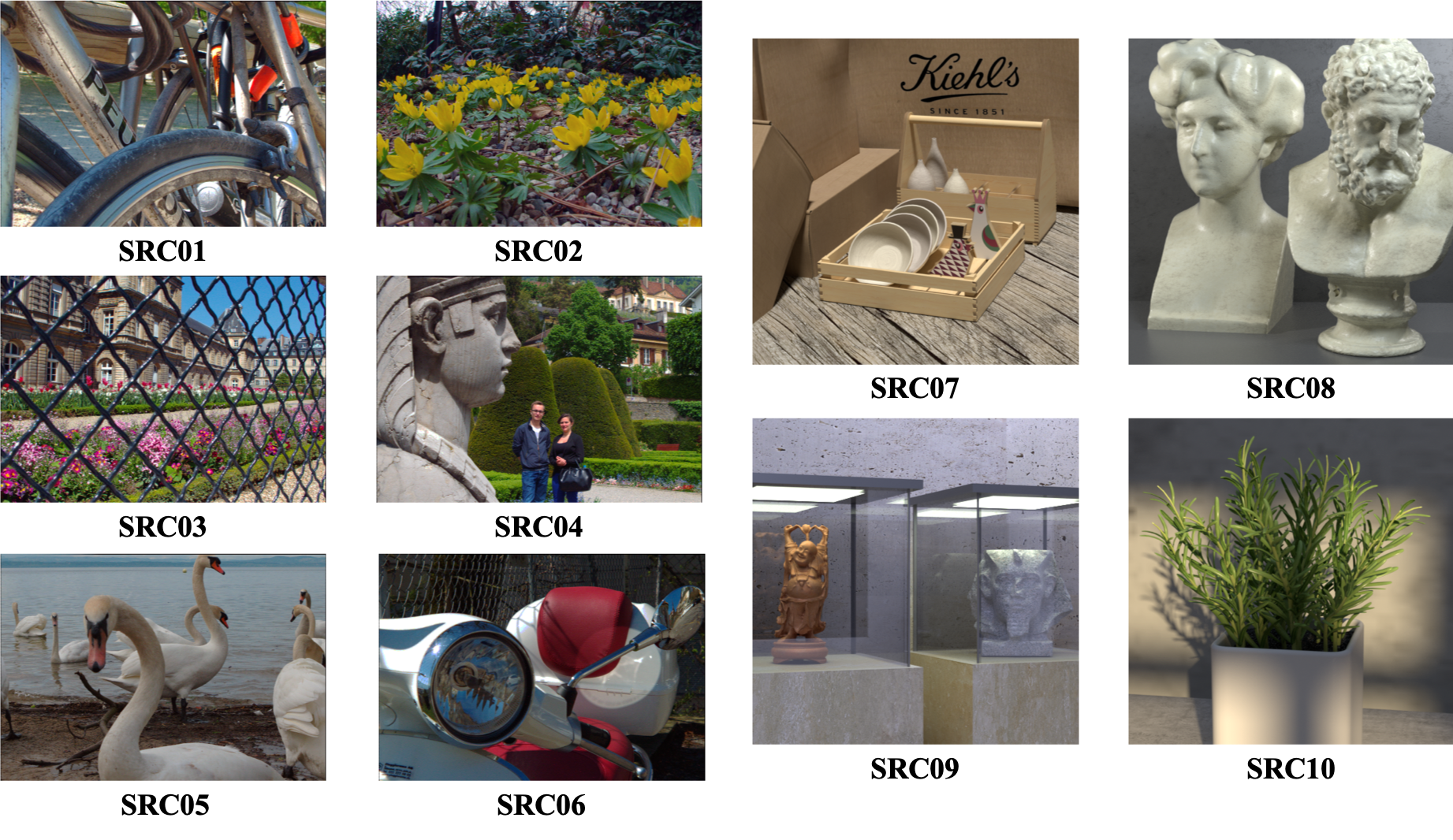}}
  \centerline{(a)}
\end{minipage}
\begin{minipage}{0.44\linewidth}
  \centerline{\includegraphics[width=4.3cm]{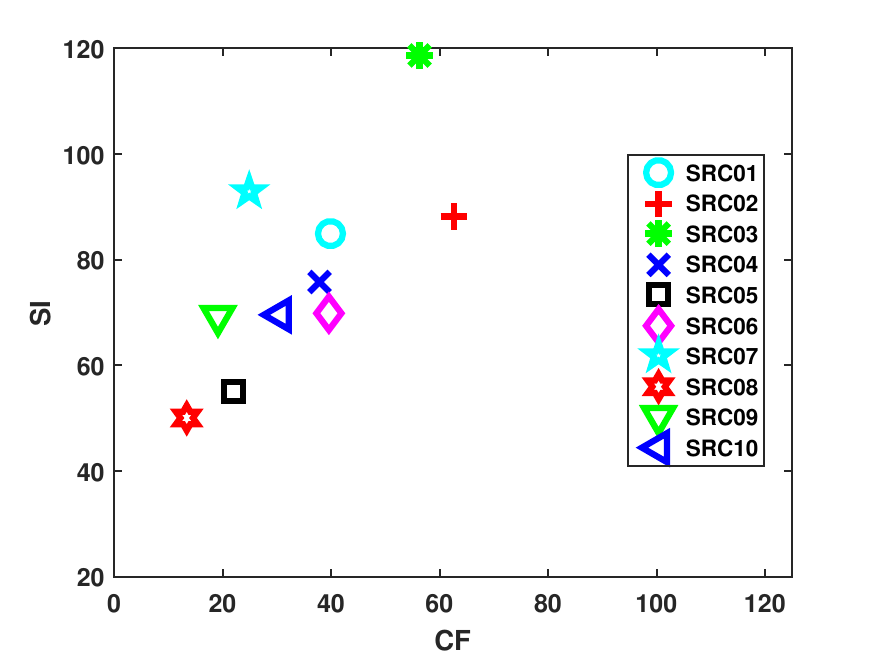}}
  \centerline{(b)}
\end{minipage}
\begin{minipage}{0.55\linewidth}
  \centerline{\includegraphics[width=4.5cm]{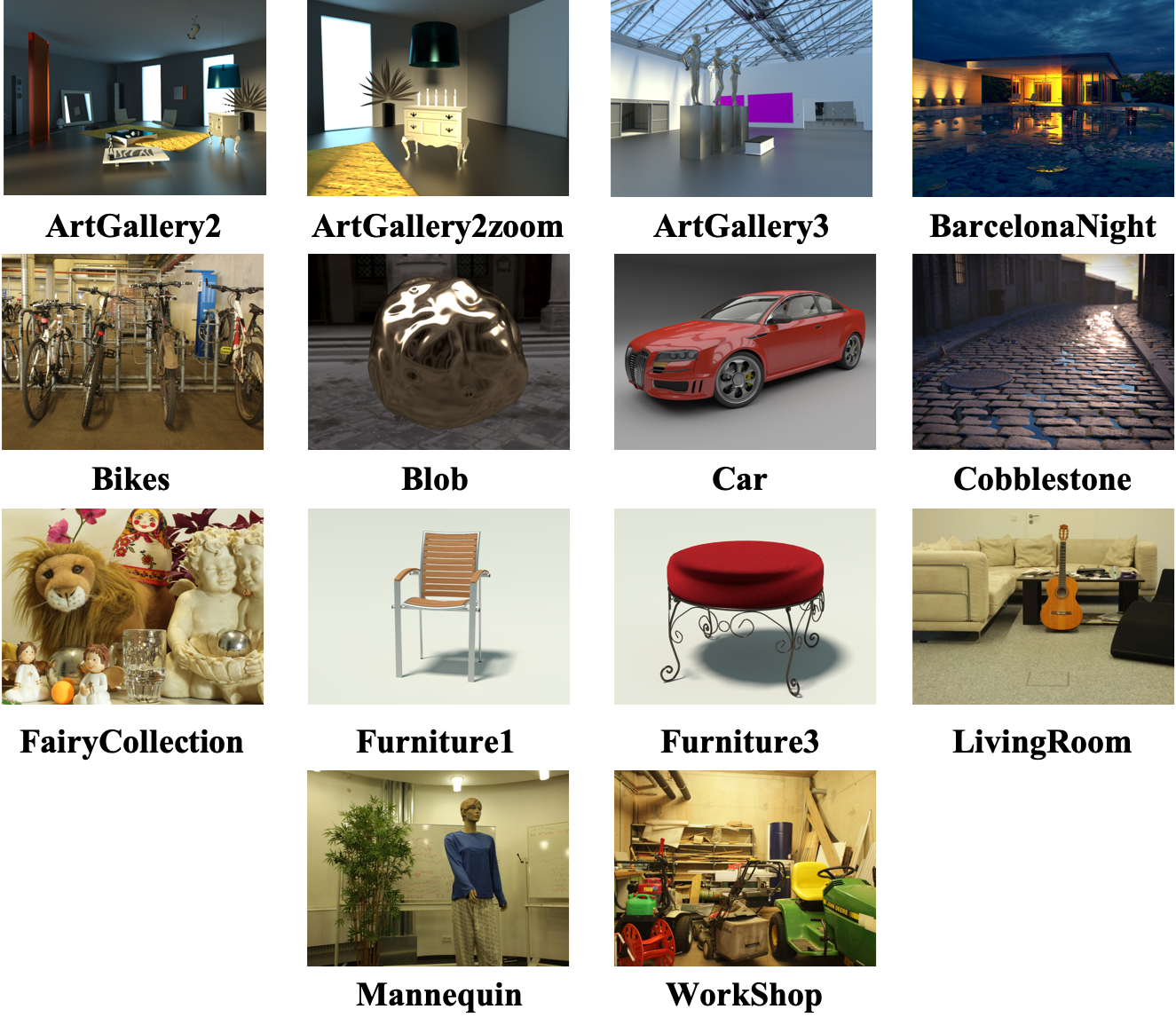}}
  \centerline{(c)}
\end{minipage}
\begin{minipage}{0.44\linewidth}
  \centerline{\includegraphics[width=4.3cm]{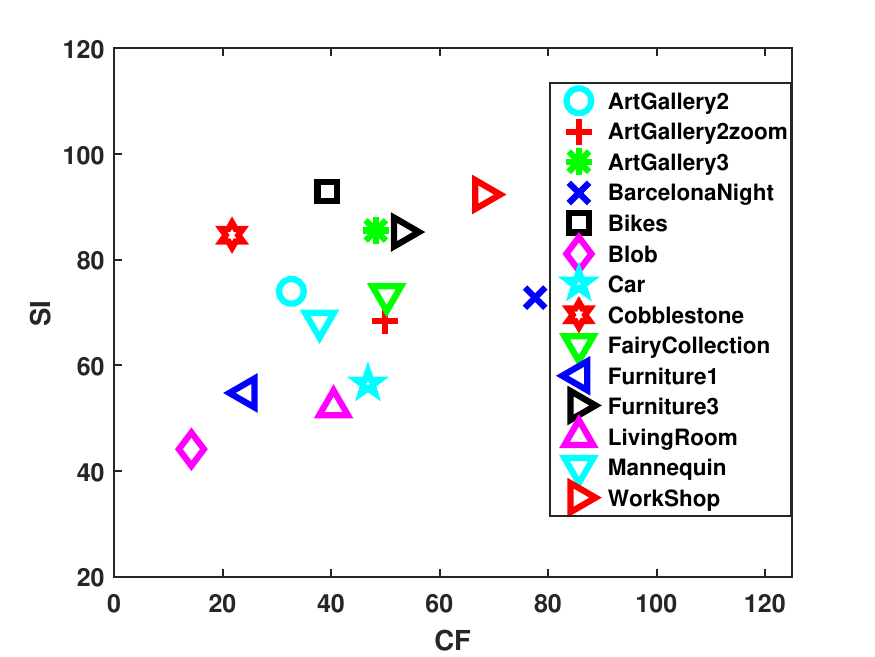}}
  \centerline{(d)}
\end{minipage}
\begin{minipage}{0.55\linewidth}
  \centerline{\includegraphics[width=4.5cm]{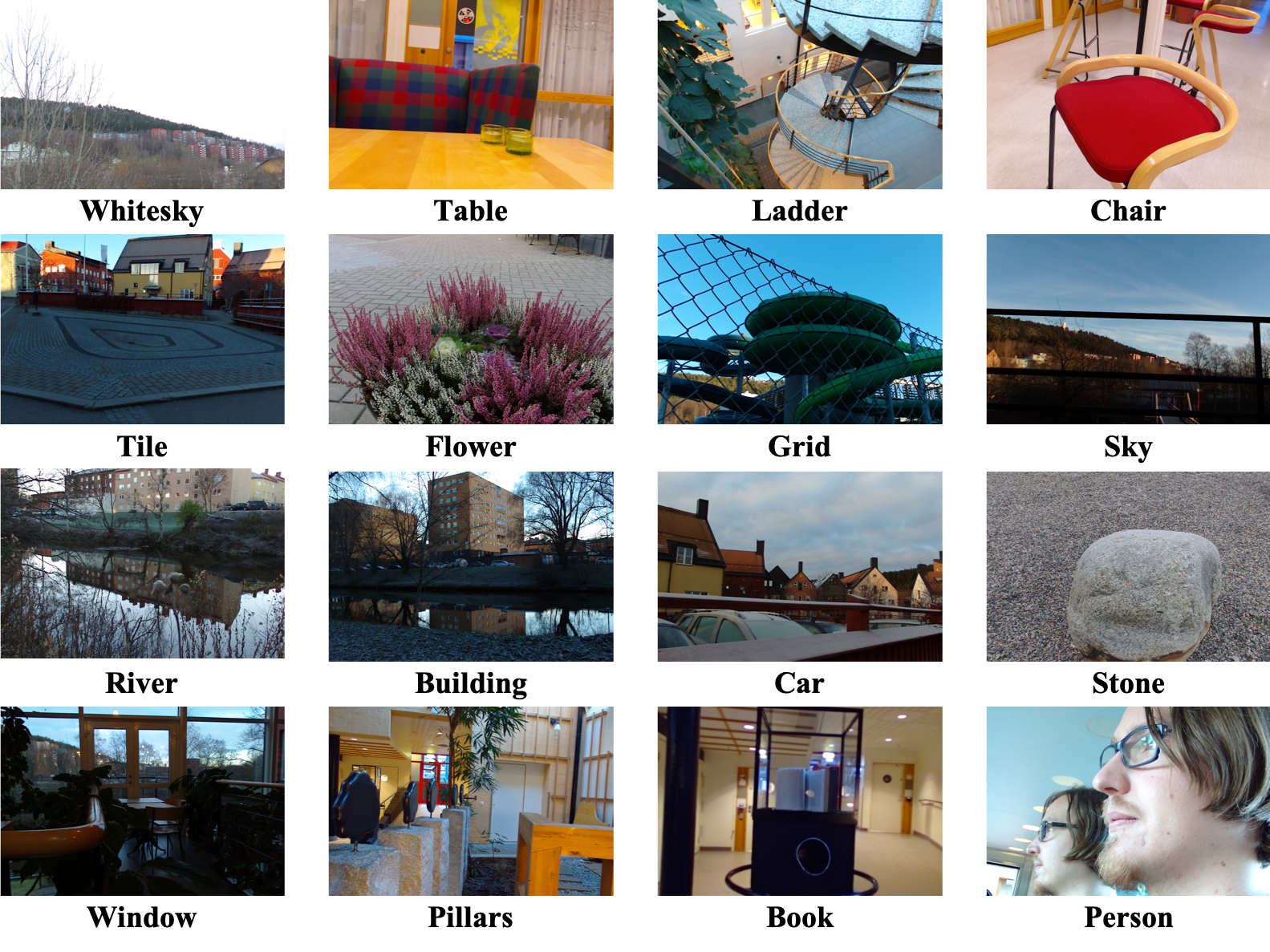}}
  \centerline{(e)}
\end{minipage}
\begin{minipage}{0.44\linewidth}
  \centerline{\includegraphics[width=4.3cm]{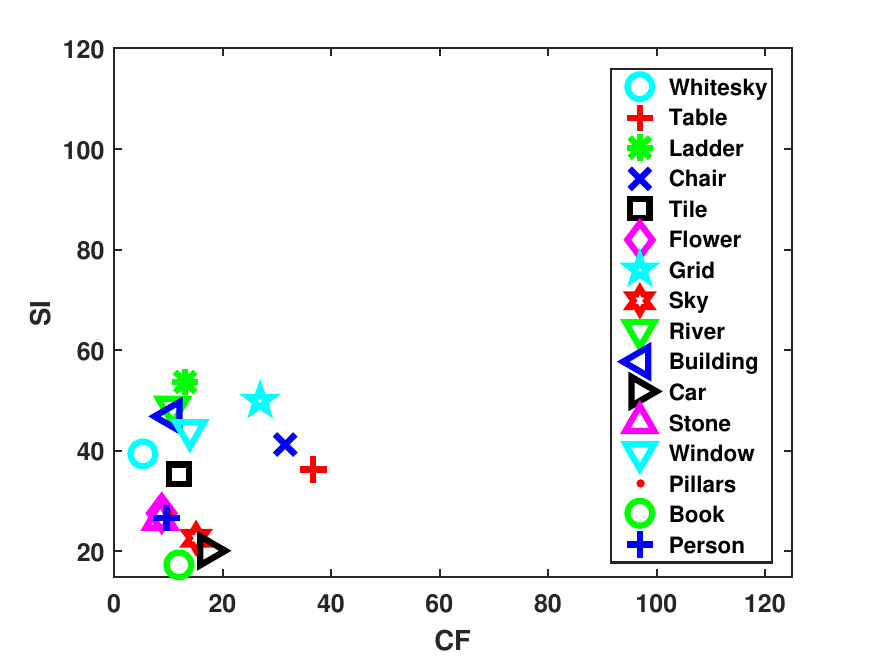}}
  \centerline{(f)}
\end{minipage}
\begin{minipage}{0.55\linewidth}
  \centerline{\includegraphics[width=4.5cm]{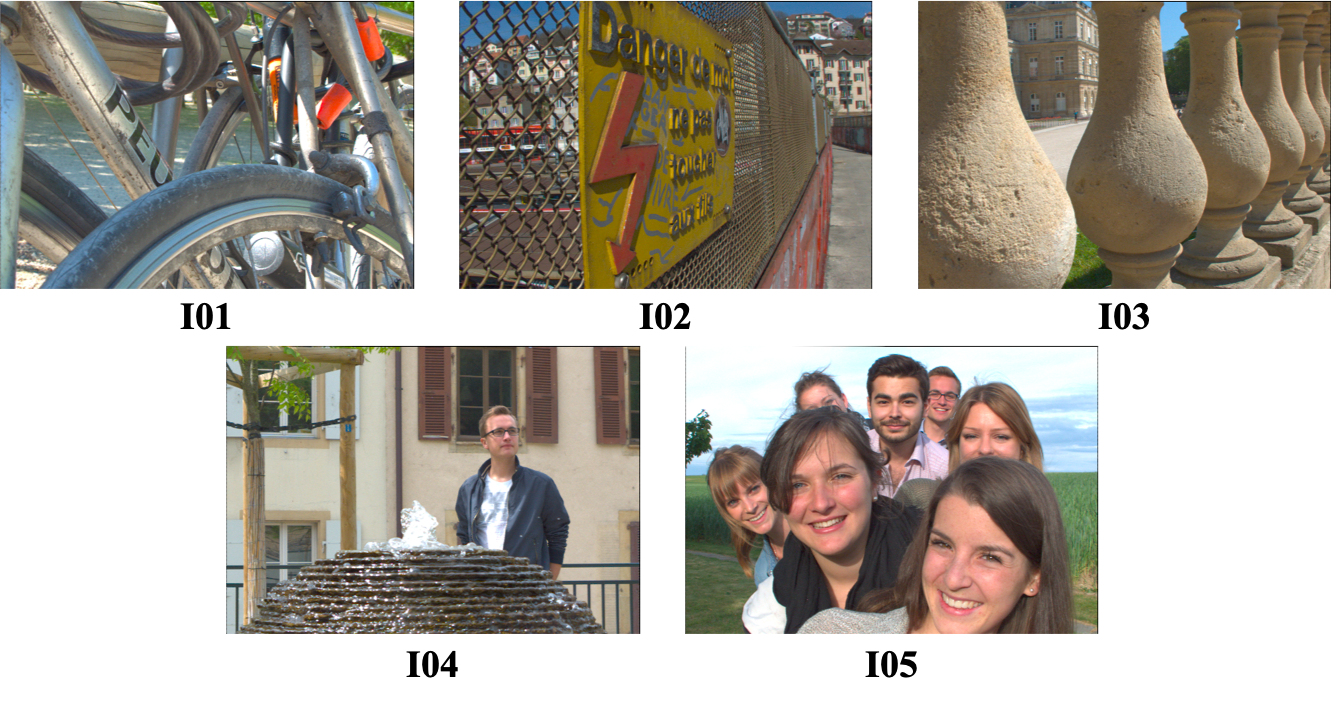}}
  \centerline{(g)}
\end{minipage}
\begin{minipage}{0.44\linewidth}
  \centerline{\includegraphics[width=4.3cm]{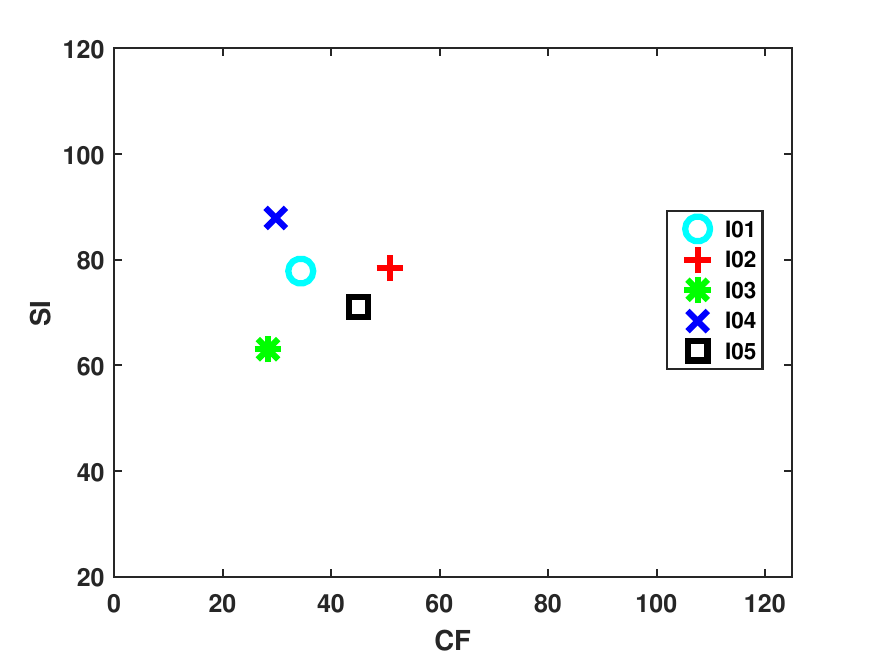}}
  \centerline{(h)}
\end{minipage}
	\caption{Database descriptions. (a) Center view of source images for Win5-LID; (b) Distribution of SI and CF for Win5-LID; (c) Center view of source images for MPI-LFA; (d) Distribution of SI and CF for MPI-LFA; (e) Center view of source images for SMART; (f) Distribution of SI and CF for SMART; (g) Center view of source images for VALID; (h) Distribution of SI and CF for VALID.}
	\label{figdatabase}
\end{figure}

\renewcommand\arraystretch{1.3}
\begin{table*}[!htb]

	\centering

	\scriptsize

	\caption{Performance Comparison on Win5-LID, MPI-LFA, and SMART Databases.}

	\begin{tabular}{c|c|cccc|cccc|cccc}

		\hline

		\multicolumn{1}{c|}{} & \multicolumn{1}{c|}{} & \multicolumn{4}{c|}{\textbf{Win5-LID}}       & \multicolumn{4}{c|}{\textbf{MPI-LFA}} & \multicolumn{4}{c}{\textbf{SMART}} \\ \hline

		\textbf{Type}                            & \textbf{Metrics}       & \textbf{SRCC}  & \textbf{LCC}   & \textbf{RMSE} & \textbf{OR}    & \textbf{SRCC}  & \textbf{LCC}   & \textbf{RMSE}  & \textbf{OR}   & \textbf{SRCC}  & \textbf{LCC}   & \textbf{RMSE} & \textbf{OR}\\ \hline

		\multirow{10}{*}{\textbf{2D FR}}         & \textbf{PSNR}          & 0.6026          & 0.6189          & 0.8031    &0.0045       & 0.8078          & 0.7830          & 1.2697  &0.0060   &0.7045 	&0.7035 	 	&1.5330 	&0.0195\\

		& \textbf{SSIM \cite{wang2004image}}          & 0.7346          & 0.7596          & 0.6650     &0.0000     & 0.7027          & 0.7123          & 1.4327 &0.0060 &0.6862 	&0.7455 	 	&1.4378 	&0.0156\\

		& \textbf{MS-SSIM \cite{wang2003multiscale}}         & 0.8266          & 0.8388          & 0.5566    &0.0000     & 0.7675          &0.7518           &1.3461  &0.0060 &0.6906 	&0.7539  	&1.4171 	&0.0117\\

		& \textbf{FSIM \cite{zhang2011fsim}}          & 0.8233          & 0.8318          & 0.5675    &0.0045     &0.7776 &0.7679 &1.3075   &0.0030   &0.7811 	&0.8139 	 	&1.2533 	&0.0039\\

		& \textbf{IW-SSIM \cite{wang2011information}}        & 0.8352          & 0.8435          & 0.5492   &0.0000   &0.8124 &0.7966 &1.2340 &0.0030     &0.7111 	&0.7971 	 	&1.3024 	&0.0000 \\

		& \textbf{IFC \cite{sheikh2005information}}           & 0.5028          & 0.5393          & 0.8611  &0.0000   &0.7573 &0.7445 &1.3629   &0.0030    &0.4827 	&0.5946 	 	&1.7343 	&0.0156\\

		& \textbf{VIF \cite{sheikh2006image}}           & 0.6665          & 0.7032          & 0.7270   &0.0000  &0.7843 &0.7861 &1.2618 &0.0030     &0.0684 	&0.2533  	&2.0867 	&0.0469   \\

		& \textbf{NQM \cite{damera2000image}}           & 0.6508          & 0.6940          & 0.7362  &0.0045   &0.7202 &0.7361 &1.3817  &0.0060   &0.4601 	&0.5305 	&1.8285 	&0.0234   \\

		& \textbf{VSNR \cite{chandler2007vsnr}}          & 0.3961          & 0.5050          & 0.8826  &0.0182  &0.7427 &0.5787 &1.6651 &0.0179  &0.5542 	&0.6289  	&1.6770 	&0.0156\\

		& \textbf{HDR-VDP2 \cite{mantiuk2011hdr}}  &0.5555 &0.6300 &0.7941 &0.0045 &0.8608 &0.8385 &1.1123 &0.0000 &0.1888 	&0.3347 	&2.0327 	&0.0625 \\ \hline

		\multirow{3}{*}{\textbf{2D NR}}          & \textbf{BRISQUE \cite{mittal2012no}}       & 0.6687          & 0.7510          & 0.5619  &0.0000   &0.6724 &0.7597 &1.1317 &0.0000    &0.8239 &0.8843 &0.8325 &0.0000 \\

		& \textbf{NIQE \cite{mittal2012making}}          & 0.2086          & 0.2645          & 0.9861   &0.0045  &0.0665 &0.1950 &2.0022   &0.0327   &0.1386 	&0.1114  	&2.1436 	&0.0547 \\

		& \textbf{FRIQUEE \cite{ghadiyaram2017perceptual}}         & 0.6328          & 0.7213          & 0.5767  &0.0000    &0.6454 &0.7451 &1.1036 &0.0000  &0.7269 &0.8345 &0.9742 &0.0000   \\ \hline

		\textbf{3D FR}                           & \textbf{Chen \cite{chen2013full}}          & 0.5269          & 0.6070          & 0.8126  &0.0091   &0.7668 &0.7585 &1.3303  &0.0030   &0.6798 	&0.7722 	&1.3706 	&0.0078  \\ \hline

		\multirow{2}{*}{\textbf{3D NR}}          & \textbf{SINQ \cite{liu2017binocular}}          & 0.8029          & 0.8362          & 0.5124  &0.0000    &0.8524 &0.8612 &0.9939 &0.0000    &0.8682 &0.8968 &0.9653 &0.0000\\

		& \textbf{BSVQE \cite{chen2018blind}}         & 0.8179          & 0.8425          & 0.4801  &0.0000     &0.8570 &0.8751 &0.9561 &0.0000  &0.8449 &0.8992 &0.8514 &0.0000  \\ \hline

		\multirow{5}{*}{\textbf{Multi-view FR}} & \textbf{MP-PSNR Full \cite{sandic2015dibr}}  & 0.5335          & 0.4766          & 0.8989   &0.0000      &0.7203 &0.6730 &1.5099     &0.0089  &0.8449 &0.8992 &0.8514 &0.0000   \\

		& \textbf{MP-PSNR Reduc \cite{sandic2016multi}} & 0.5374          & 0.4765          & 0.8989   &0.0000    &0.7210 &0.6747 &1.5067  &0.0089   &0.6716 	&0.6926 	&1.5559 	&0.0117  \\

		& \textbf{MW-PSNR Full \cite{sandic2015dibr1}}  & 0.5147          & 0.4758          & 0.8993   &0.0000      &0.7232 &0.6770 &1.5023   &0.0089   &0.6620 	&0.6505 	&1.6382 	&0.0117 \\

		& \textbf{MW-PSNR Reduc \cite{sandic2015dibr1}} & 0.5326          & 0.4766          & 0.8989   &0.0000    &0.7217 &0.6757 &1.5048  &0.0089   &0.6769 	&0.6903 	&1.5607 	&0.0117  \\

		& \textbf{3DSwIM \cite{battisti2015objective}}        & 0.4320          & 0.5262          & 0.8695   &0.0182    &0.5565 &0.5489 &1.7063 &0.0119  &0.4053 	&0.4707 	&1.9032 	&0.0234 \\ \hline

		\textbf{Multi-view NR}                  & \textbf{APT \cite{gu2018model}}           & 0.3058          & 0.4087          & 0.9332  &0.0045   &0.0710 &0.0031 &2.0413 &0.0357  &0.5105   & 0.5249 &1.8361 &0.0234\\ \hline

		\textbf{LFI RR}                  & \textbf{LF-IQM \cite{8632960}} & 0.4503 & 0.4763 & 0.8991 & 0.0273 &0.3364 &0.4223 &1.8504 & 0.0268 &0.1222 &0.2998 &2.0579 &0.0547 \\ \hline
		\multirow{2}{*}{\textbf{LFI NR}} &\textbf{BELIF \cite{shi2019belif}}  &0.8719 &0.8910 &0.4294 &0.0000 &0.8854 &0.9096 &0.7877 &0.0000 &0.8367 &0.8833 &0.8347 &0.0000 \\
		& \textbf{Proposed Tensor-NLFQ} & \textbf{0.9101} & \textbf{0.9217} & \textbf{0.3781} & \textbf{0.0000} & \textbf{0.9221} & \textbf{0.9294} & \textbf{0.7241} & \textbf{0.0000} &\textbf{0.8702} &\textbf{0.9028} &\textbf{0.8225} &\textbf{0.0000} \\ \hline

	\end{tabular}
\label{table_overall}
\end{table*}

\renewcommand\arraystretch{1.3}
\begin{table*}[!htb]
	\centering
	\scriptsize
	\caption{Performance Comparison on VALID Database.}
	\begin{tabular}{c|c|cccc|cccc}
		\hline
		\multicolumn{1}{c|}{} & \multicolumn{1}{c|}{} & \multicolumn{4}{c|}{\textbf{VALID-8bit }}  & \multicolumn{4}{c}{\textbf{VALID-10bit }}\\ \hline
		\textbf{Type}                            & \textbf{Metrics}       & \textbf{SRCC}  & \textbf{LCC}   & \textbf{RMSE} & \textbf{OR}    & \textbf{SRCC}  & \textbf{LCC}   & \textbf{RMSE}  & \textbf{OR}  \\ \hline
		\multirow{6}{*}{\textbf{2D FR}}         & \textbf{PSNR}         &0.9620 &0.9681 &0.3352 &0.0000  &0.9467  &0.9524  &0.2935 &0.0000\\
		& \textbf{SSIM \cite{wang2004image}}           & 0.9576          & 0.9573          & 0.3868 &0.0000  &0.9326 &0.9375 &0.3348 &0.0000\\
		& \textbf{MS-SSIM \cite{wang2003multiscale}}         &0.9593 &0.9658   &0.3473 &0.0000  &0.9432 &0.9484 &0.3051 &0.0000\\
		& \textbf{IW-SSIM \cite{wang2011information}}        &0.9674 &0.9764 &0.2892   &0.0000 &\textbf{0.9499} &0.9617 &0.2638 &0.0000 \\
		& \textbf{NQM \cite{damera2000image}}              &0.9055 &0.9194 &0.5266   &0.0000  &0.8410 &0.8582 &0.4940 &0.0000\\
		& \textbf{HDR-VDP2 \cite{mantiuk2011hdr}}  &0.9623 &0.9785 &0.2758 &0.0000 &0.9371 &0.9528 &0.2921 &0.0000 \\ \hline
		\multirow{2}{*}{\textbf{2D NR}}          & \textbf{BRISQUE \cite{mittal2012no}}       &0.9222 &0.9849 &0.2017   &0.0000 &0.9027 &0.9347 &0.2838 &0.0000 \\
		& \textbf{FRIQUEE \cite{ghadiyaram2017perceptual}}        &0.9157 &0.9836 &0.2160   &0.0000 &0.8559 &0.8986 &0.3497 &0.0000  \\ \hline
		\textbf{3D NR}          & \textbf{SINQ \cite{liu2017binocular}}         &0.9222 &0.9849 &0.2070    &0.0000  &0.9021 &0.9348 &0.2722 &0.0000 \\
		\multirow{5}{*}{\textbf{Multi-view FR}} & \textbf{MP-PSNR Full \cite{sandic2015dibr}}   &0.9730 &0.9852 &0.2291  &0.0000  &0.3830 &0.3582 &0.8986 &0.0000\\
		& \textbf{MP-PSNR Reduc \cite{sandic2016multi}}  &\textbf{0.9744} &0.9859 &0.2237   &0.0000 & 0.3826 &0.3506 &0.9013 &0.0000\\
		& \textbf{MW-PSNR Full \cite{sandic2015dibr1}}     &0.9597 &0.9677 &0.3376   &0.0000  &0.3764 &0.3556 &0.8995 &0.0000\\
		& \textbf{MW-PSNR Reduc \cite{sandic2015dibr1}} &0.9648 &0.9751 &0.2970   &0.0000  &0.3815 &0.3563 &0.8993 &0.0100\\
		& \textbf{3DSwIM \cite{battisti2015objective}}         &0.7950 &0.7876 &0.8248  &0.0000 &0.7869 &0.7401 &0.6472 &0.0000\\ \hline
		\textbf{LFI RR}                  & \textbf{LF-IQM \cite{8632960}}             &0.3934   &0.5001   &1.1593    &0.0000 &0.3679 &0.3705 &0.8939 &0.0100 \\ \hline
		\multirow{2}{*}{\textbf{LFI NR}} &\textbf{BELIF \cite{shi2019belif}}  &0.9278 &\textbf{0.9862} &\textbf{0.1680} &0.0000 &0.9186 &0.9622 &0.2387 &0.0000 \\
		& \textbf{Proposed Tensor-NLFQ} & 0.9286 & 0.9852 & 0.1825 & \textbf{0.0000} & 0.9367 & \textbf{0.9640} & \textbf{0.2295} & \textbf{0.0000} \\ \hline
	\end{tabular}
\label{table_valid}
\end{table*}

In addition to spatial quality, angular consistency also affects the LFI quality. Usually, angular reconstruction operations, such as interpolation, may break angular consistency. To measure the degradation of angular consistency, we propose the tensor angular variation index. Specifically, we first compute the structural similarity between each view in the view stack and its corresponding first principal component:

\begin{equation}
\textbf{ss}_{n}^{d}(i) = F(C_{n}^{d}(i), M_n^{d}),
\end{equation}
where $C_{n}^{d}$ is the input view stack and $M_n^{d}$ represents the corresponding first principal component. $i$ indicates the angular coordinate of $C$. $n = 1, 2, 3$ and $d = {0^{\circ}, 45^{\circ}, 90^{\circ}, 135^{\circ}}$ represent the index of three color channels and four orientations, respectively. $F$ is the function to calculate the structural similarity between $C_{n}^{d}(i)$ and $M_n^{d}$. In our paper, we use the SSIM \cite{wang2004image}.

\begin{figure}[!htb]
	\centering
	\begin{minipage}{0.47\linewidth}
		\centerline{\includegraphics[width=4.5cm]{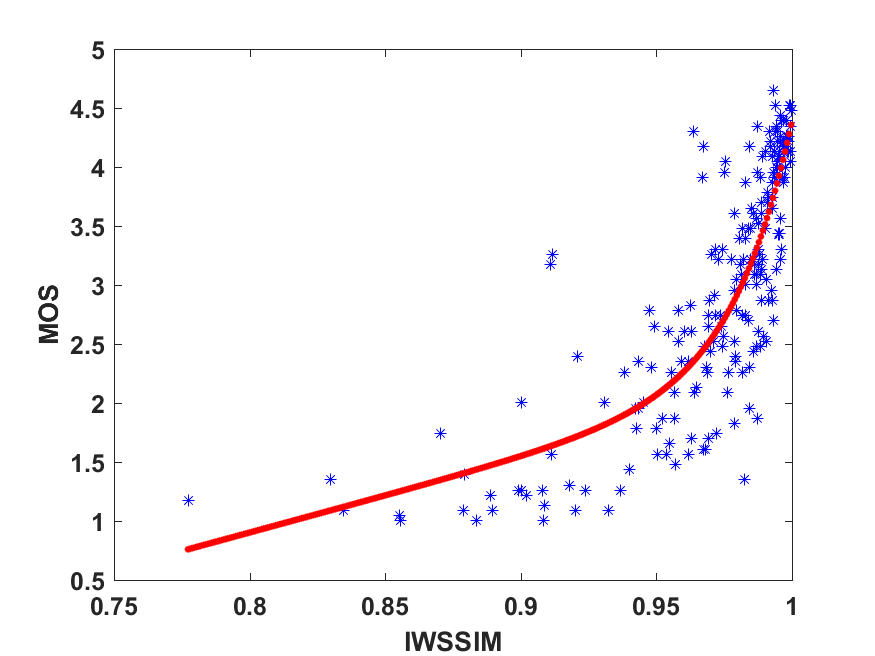}}
		\centerline{\includegraphics[width=4.5cm]{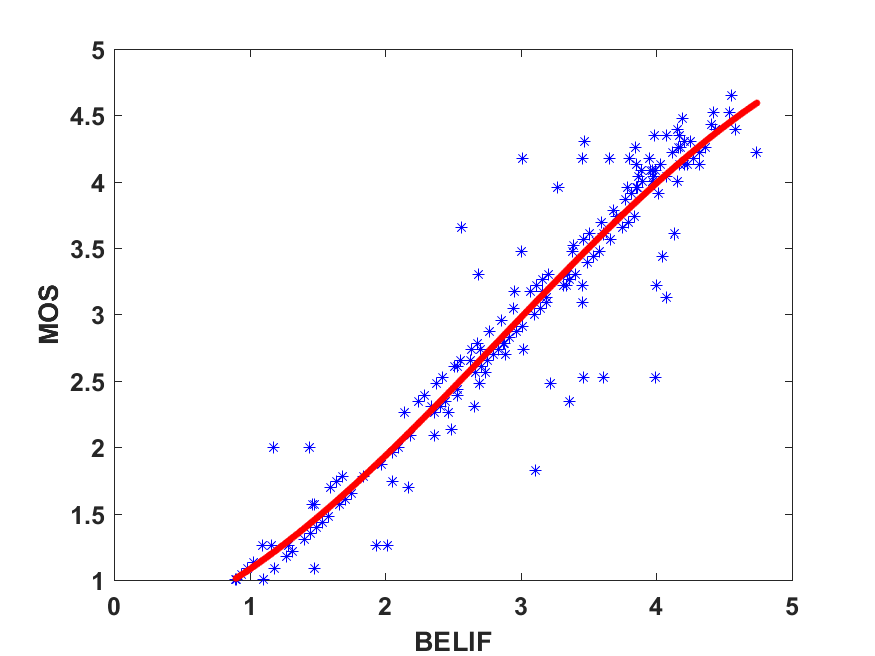}}
		\centerline{\includegraphics[width=4.5cm]{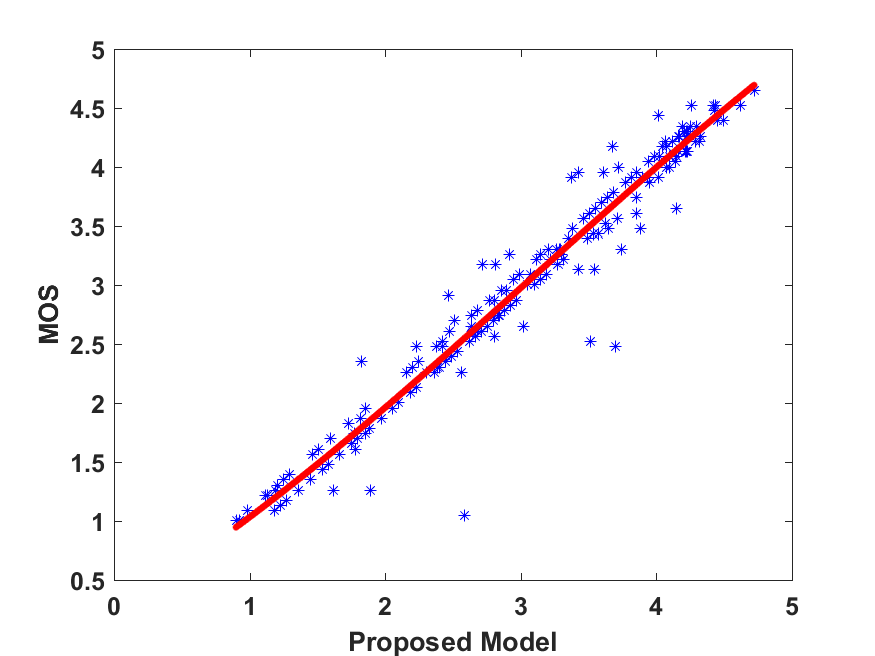}}
		\centerline{(a)}
	\end{minipage}
	\begin{minipage}{0.47\linewidth}
		\centerline{\includegraphics[width=4.5cm]{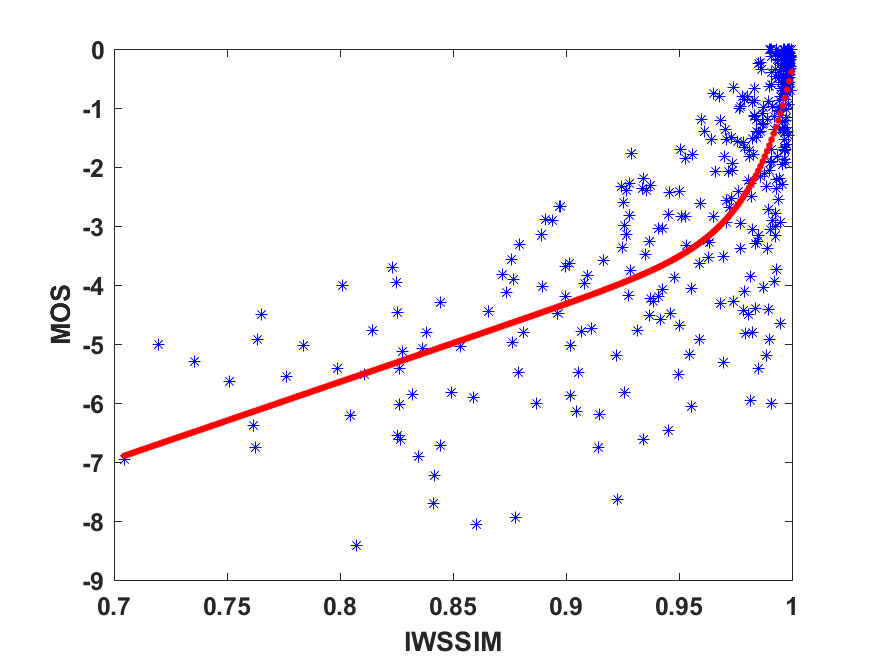}}
		\centerline{\includegraphics[width=4.5cm]{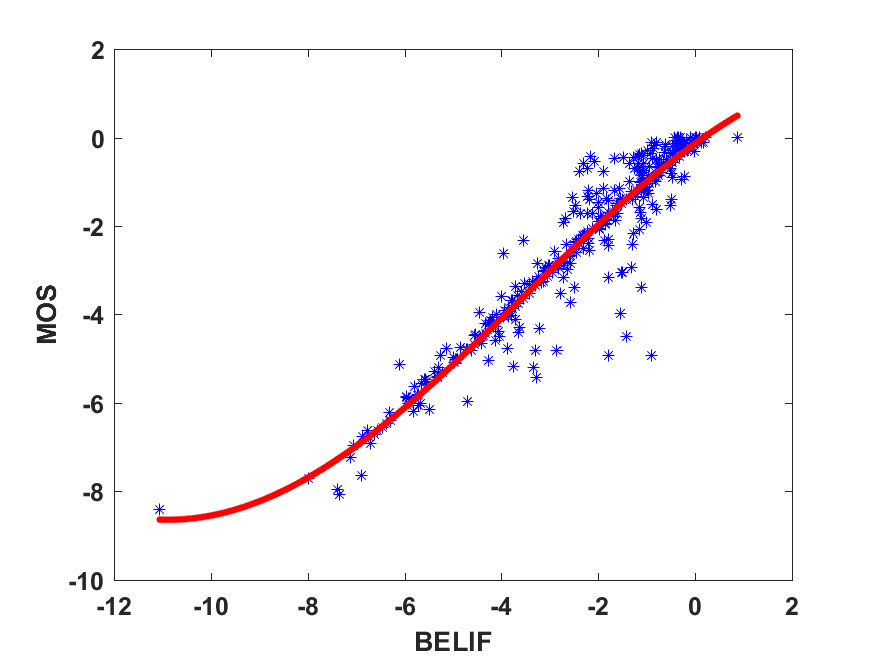}}
		\centerline{\includegraphics[width=4.5cm]{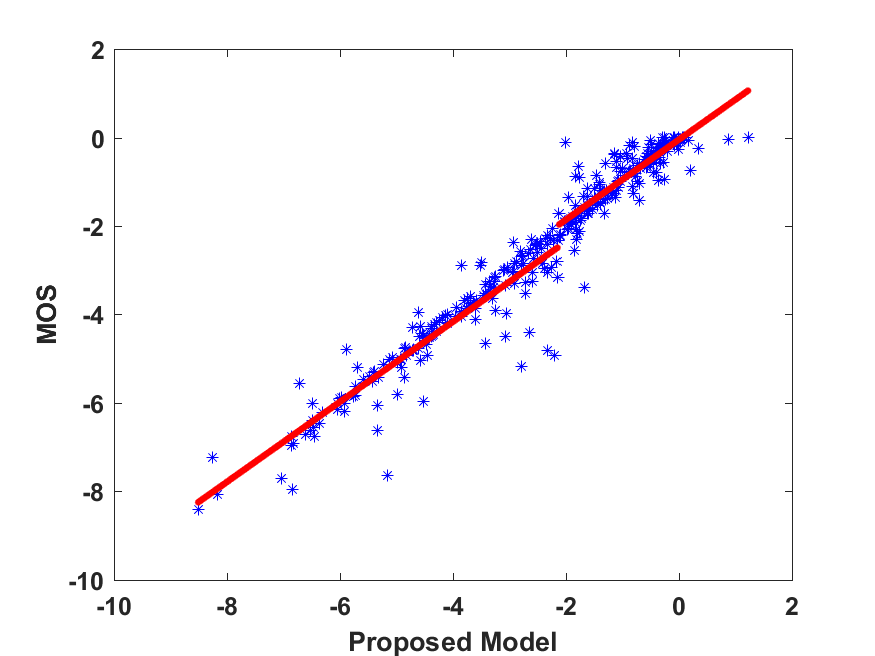}}
		\centerline{(b)}
	\end{minipage}
	\caption{Scatter plots of predicted quality scores by three methods against the MOS values on the Win5-LID and MPI-LFA databases. The horizontal and vertical axes in each figure represent the predicted quality scores and MOS values, respectively. The red line is the fitted curve. The top, middle and bottom rows are the results of IW-SSIM, BELIF and the proposed model, respectively. (a) Scatter plots on Win5-LID database; (b) Scatter plots on MPI-LFA database.}
	\label{figscatter}
\end{figure}

The structure similarity distribution of LFI selected from MPI-LFA \cite{adhikarla2017towards} is illustrated in Fig. \ref{figSS}. For information with only horizontal direction, we only use horizontal direction features and the weight of other directions feature is 0. As shown in Fig. \ref{figSS}, the horizontal view stack can also reflect the change of angular consistency. In other words, since the MPI-LFA database contains only horizontal LFIs, Fig. \ref{figSS} only presents the structure similarity distribution of the horizontal view stack. The top row represents the $\textbf{ss}_{1}^{0^\circ}$ of luminance $L$ channel, while the middle and bottom rows indicate the $\textbf{ss}_{2}^{0^\circ}$ and $\textbf{ss}_{3}^{0^\circ}$ of chrominance $a$ and $b$ channels, respectively. Fig. \ref{figSS}(a-d) show the structure similarity distribution of original LFI and the distribution of different distortion levels for quantized depth maps (DQ), linear interpolation (LINEAR), nearest interpolation (NN), and image warping using optical flow estimation (OPT) artifacts.
It can be seen that when the angular consistency is not destroyed, the distribution of structural similarity is smooth, as shown the cyan curve in Fig. \ref{figSS}. However, when the angular consistency is degraded by interpolation distortion, the distribution of structural similarity changes significantly. Specifically, as the angular consistency deteriorates, the variation degree in the structural similarity distribution of the LFI increases gradually. Moreover, different distortions types have different wave shapes. For example, the NN distortion is stepped and the LINEAR distortion has more peaks. These demonstrate that the structure similarity distribution is good at distinguishing various distortion types and levels.

Then, inspired by Fig. \ref{figSS}, we employ a second-order polynomial to fit the structure similarity distribution as follows:

\begin{equation}
\centering
 \textbf{ss}_n^d(i) = f_1 i^2 + f_2 i + f_3,
\end{equation}
where $i$ is the angular coordinate. $f_1$, $f_2$ and $f_3$ are fitting parameters modeling variation of angular consistency.

To further characterize the structure similarity properties, we extract several complementary features including contrast, angular second moment, entropy and inverse different moment \cite{kim2015quantitative } to represent the deterioration information. Specifically, the contrast is the amount of local variation presented in structure similarity. The angular second moment and inverse different moment measure the homogeneity. Thus, $\textbf{f}_{TAVI}$ is obtained by concatenating the fitting parameters (i.e. $f_1$, $f_2$, $f_3$) and the complementary features. The dimension of feature TAVI is 30.

\subsubsection{Direction Pooling}
For a LFI with an angular resolution of $S\times T$, we have $S$ horizontal view stacks, $T$ vertical view stacks, $S+T-1$ main-diagonal view stacks and $S+T-1$ secondary-diagonal view stacks. Since we extract the features of the view stack in each orientation and average the features from the stack in the same orientation, the $\textbf{f}_{d}$ is first obtained by concatenating $\textbf{f}_{PCSC}$ and $\textbf{f}_{TAVI}$ in the same orientation and $d = {0^{\circ}, 45^{\circ}, 90^{\circ}, 135^{\circ}}$. We then model the final features by:

\begin{equation}
\centering
\textbf{f}_{final} = w_1 \textbf{f}_{0^{\circ}} + w_2 \textbf{f}_{45^{\circ}} + w_3 \textbf{f}_{90^{\circ}} +  w_4 \textbf{f}_{135^{\circ}} ,
\end{equation}
where $w_1$, $w_2$, $w_3$ and $w_4$ indicate the corresponding weights of four orientations. In our model, we set $w_1 = w_2 = w_3 = w_4 = \frac{1}{4}$. The final results are trained with the average features from each stack in the same orientation, and then all directions are weighted.

Finally, in this model, we train a regression model to map the final feature vector $\textbf{f}_{final}$ space to quality scores. In our implementation, we adopt the well-known support vector regression (SVR), which has been effectively applied to many image quality assessment problems \cite{zhou20163d,chen2018blind,liu2017binocular}. Specifically, the LIBSVM package \cite{chang2011libsvm} is utilized to implement the SVR with a radial basis function kernel.


\section{Experimental Results}
\begin{figure}
	\centering
	\begin{minipage}{0.47\linewidth}
		\centerline{\includegraphics[width=4.5cm]{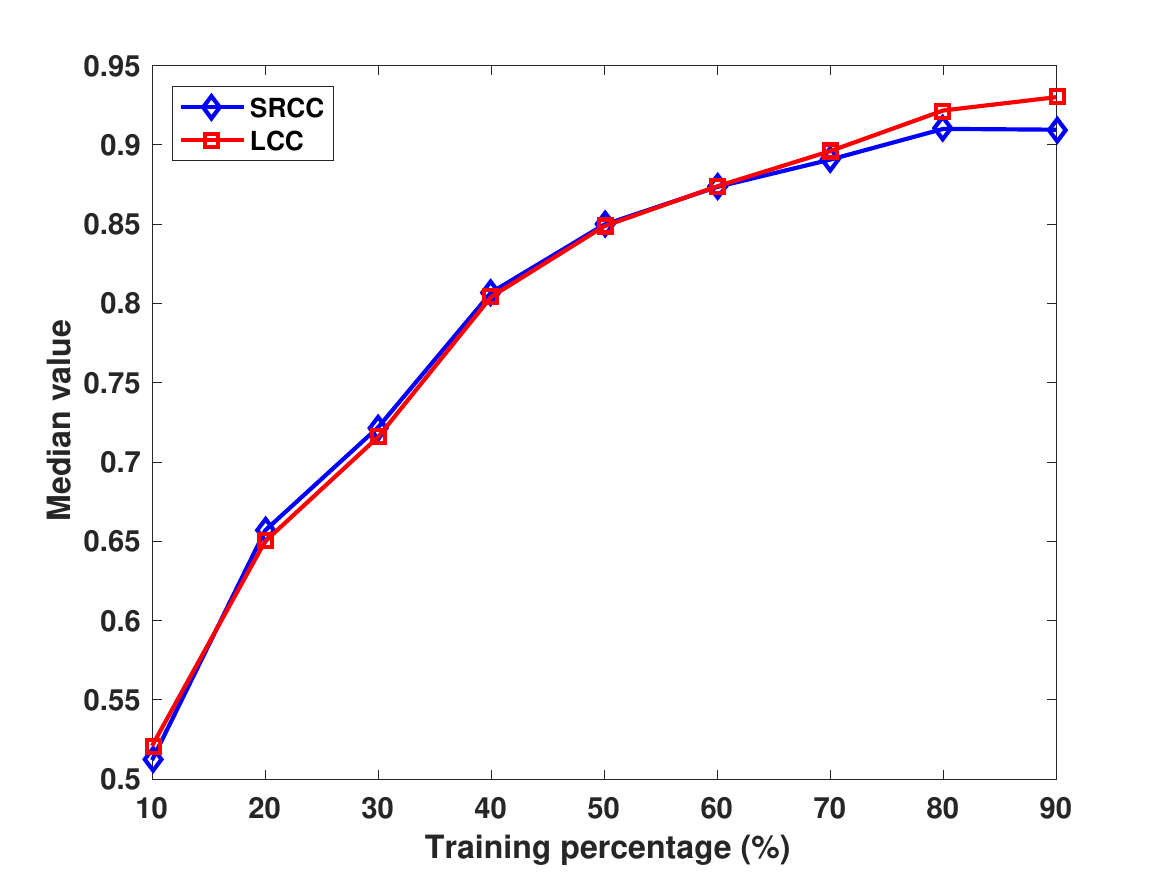}}
		\centerline{(a)}
	\end{minipage}
	\begin{minipage}{0.47\linewidth}
		\centerline{\includegraphics[width=4.5cm]{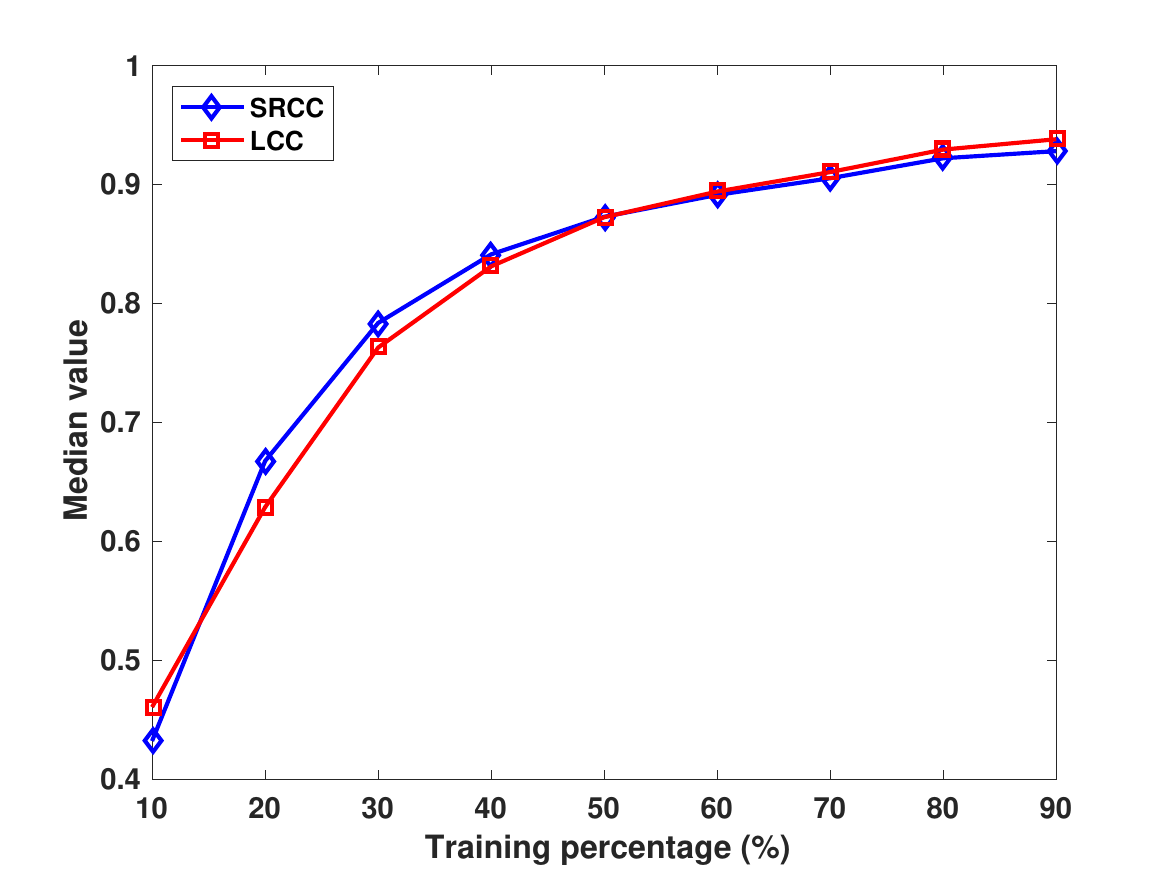}}
		\centerline{(b)}
	\end{minipage}
	\caption{The change of performance results for Tensor-NLFQ with different training and testing percentages. (a) Run on Win5-LID database; (b) Run on MPI-LFA database.}
	\label{figpercentages}
\end{figure}

\renewcommand\arraystretch{1.3}
\begin{table*}[!htb]
	\scriptsize
	\centering
	\caption{RMSE Performance of Different Distortion Types on Win5-LID and MPI-LFA Databases.}
	\begin{tabular}{c|c|cccc|cccccc}
		\hline
&		& \multicolumn{4}{c|}{\textbf{Win5-LID}}       & \multicolumn{6}{c}{\textbf{MPI-LFA}}   \\ \hline
\textbf{Type}		&\textbf{Metrics} & \textbf{HEVC}   & \textbf{JPEG}   & \textbf{LINEAR}     & \textbf{NN}   & \textbf{HEVC}   & \textbf{DQ}     & \textbf{OPT}    & \textbf{LINEAR} & \textbf{NN}     & \textbf{GAUSS}  \\ \hline
\multirow{10}{*}{\textbf{2D FR}}		&\textbf{PSNR}    & 0.6404          & 0.5219          & 0.5233          & 0.3843        & 0.9496          & 1.1390          & 1.3430          & 0.9156          & 0.7667          & 0.7894  \\
		&\textbf{SSIM \cite{wang2004image}}    & 0.3808          & 0.5324          & 0.4414          & 0.4252        & 0.5578          & 1.5566          & 1.4087          & 1.2695          & 1.0692          & 0.4188  \\
		&\textbf{MS-SSIM \cite{wang2003multiscale}}   & 0.2515          & 0.4272          & 0.4167          & 0.4170      & 0.4488          & 1.2445          & 1.3018          & 1.0766          & 0.7961          & 0.5049    \\
		&\textbf{FSIM \cite{zhang2011fsim}}    & 0.2564          & 0.4377          & 0.4062          & 0.4009      & 0.5091          & 1.0886          & 1.2849          & 1.0537          & 0.8142          & 0.6747    \\
		&\textbf{IW-SSIM \cite{wang2011information}}  & 0.2559          & 0.4133          & 0.3702          & 0.4187       & 0.4936          & 1.0900          & 1.1853          & 0.8957          & 0.6105          & 0.5343   \\
		&\textbf{IFC \cite{sheikh2005information}}     & 0.5396          & 0.6433          & 0.7092          & 0.4472     & 0.8244          & 0.8882          & 1.1915          & 0.8121          & 0.6229          & 0.9893     \\
		&\textbf{VIF \cite{sheikh2006image}}     & 0.2792          & 0.4351          & 0.4554          & 0.3712      & 0.4336          & 1.2041          & 1.4072          & 0.8385          & 0.7069          & 0.6535    \\
		&\textbf{NQM \cite{damera2000image}}     & 0.4952          & 0.6062          & 0.6228          & 0.5248     & 0.6618          & 1.3064          & 1.4238          & 1.1291          & 0.6228          & 0.7660      \\
		&\textbf{VSNR \cite{chandler2007vsnr}}    & 0.4460          & 0.4585          & 0.9115          & 0.7314       & 0.3249          & 1.7150          & 1.5878          & 1.6959          & 1.3126          & 0.6930   \\
& \textbf{HDR-VDP2 \cite{mantiuk2011hdr}}& 0.4460          & 0.4585          & 0.9115          & 0.7314       & 0.3249          & 1.7150          & 1.5878          & 1.6959          & 1.3126          & 0.6930   \\ \hline
\multirow{3}{*}{\textbf{2D NR}}		&\textbf{BRISQUE \cite{mittal2012no}} & 0.2695          & 0.2225          & 0.3230          & 0.4018     &0.3468          & 0.7297          & 0.8132          & 0.7577          &  0.7733          & 0.4301     \\
		&\textbf{NIQE \cite{mittal2012making}}    & 0.9978          & 0.9686          & 0.8986          & 0.7594      & 1.0622          & 1.7489          & 1.7262          & 2.0515          & 1.6397          & 1.9197      \\
		&\textbf{FRIQUEE \cite{ghadiyaram2017perceptual}}   & 0.4127          & 0.2273          & 0.3285          & 0.3748     & 0.2605          & 0.6906          & 0.8170          & 0.8067          & 0.8002          & 0.2551     \\ \hline
\textbf{3D FR}		&\textbf{Chen \cite{chen2013full}}   & \textbf{0.2341} & 0.3795          & 0.8978          & 0.6077        & 0.4763          & 1.2547          & 1.3122          & 1.0762          & 0.8093          & 0.4824   \\ \hline
\multirow{2}{*}{\textbf{3D NR}}		&\textbf{SINQ \cite{liu2017binocular}}    & 0.2373          & 0.2169          & 0.2846          & 0.3419      & \textbf{0.1638}          & 0.5996          & 0.8109          & 0.6511          & 0.7078          & 0.3432      \\
		&\textbf{BSVQE \cite{chen2018blind}}   & 0.2943          & 0.2300          & 0.3199         & 0.2628       & 0.3128          & 0.6877          & 0.8255          & 0.6571          & 0.4864          & 0.2309    \\ \hline
\multirow{5}{*}{\textbf{Multi-view FR}}		&\textbf{MP-PSNR Full \cite{sandic2015dibr}}  & 0.5357          & 0.5216          & 0.7325    &0.4318      & 0.7634          & 0.9989          & 1.1191  	&0.7106   &0.6670     &0.9100    \\
		&\textbf{MP-PSNR Reduc \cite{sandic2016multi}} & 0.4592          & 0.5119          & 0.7331     &0.4255     & 0.9392          & 0.9998          & 1.1181  &0.7179   &0.6608     &0.9201     \\
		&\textbf{MW-PSNR Full \cite{sandic2015dibr1}}  & 0.5857          & 0.5511          & 0.7329    &0.4291       & 0.8729          & 1.0013          & 1.1155  &0.6957   &0.6426     &0.7260   \\
		&\textbf{MW-PSNR Reduc \cite{sandic2015dibr1}} & 0.4762          & 0.5326          & 0.7329    &0.4313        & 0.8056          & 0.9980          & 1.1135  &0.7130   &0.6519     &0.7572  \\
		&\textbf{3DSwIM \cite{battisti2015objective}}        & 0.9778          & 0.7278          & 0.6160  &0.4536     & 1.9656          & 1.2155          & 1.4833  &1.2782   &1.2013     &1.5171    \\ \hline
\textbf{Multi-view NR}		&\textbf{APT \cite{gu2018model}}   & 0.9788  & 0.9757          & 0.7731          & 0.7196       & 2.2549          & 1.7238          & 1.6306  &2.0857   &1.6486     &1.9675   \\ \hline
\textbf{LFI RR}		&\textbf{LF-IQM \cite{8632960}}           & 1.0987          & 0.8029          & 0.6096   &0.5409  &1.8225    &1.7211	   &1.7645   &1.9623   &1.5168     &1.3558      \\ \hline
\multirow{2}{*}{\textbf{LFI NR}}		&\textbf{BELIF \cite{shi2019belif}}     & 0.3062          & 0.2013 & 0.3187 & 0.3052 &0.2486 & 0.6513 & 0.8023 & 0.4472 & 0.3556 & \textbf{0.1527}    \\
		&\textbf{Proposed Tensor-NLFQ}     & 0.2578          & \textbf{0.1902} & \textbf{0.2633} & \textbf{0.1974} &0.2030 & \textbf{0.5370} & \textbf{0.7970} & \textbf{0.3555} & \textbf{0.3125} & 0.1597    \\ \hline
	\end{tabular}
	\label{tbsingledis}
\end{table*}

To validate our proposed Tensor-NLFQ model, we conduct experiments on four publicly available databases, namely Win5-LID \cite{shi2018light}, MPI-LFA \cite{adhikarla2017towards}, SMART \cite{paudyal2017towards} and VALID \cite{viola2018valid}.

As shown in Fig. \ref{figdatabase}(a) and Fig. \ref{figdatabase}(b), the Win5-LID database contains 6 real scenes captured by Lytro illum and 4 synthetic scenes as original images which cover various spatial perceptual information (SI) and colorfulness (CF) \cite{itu1999subjective}. There exist 220 distorted LFIs by introducing 6 distortion types, including HEVC, JPEG, LINEAR, NN and two CNN models. Moreover, more than 20 observers are invited to provide subjective ratings for the 220 distorted LFIs under the double-stimulus continuous quality scale on a 5-point discrete scale. Therefore, each LFI has the overall mean opinion score (MOS) value, which is a statistical concept as the ground truth image quality measurement. The MOS is calculated by the mean subjective ratings of each LFI.

As we can see from Fig. \ref{figdatabase}(c) and Fig. \ref{figdatabase}(d), the MPI-LFA database consists of 14 pristine LFIs captured by the TSC system, which also cover various SI and CF. The 336 distorted LFIs are produced with 6 distortion types, i.e. HEVC, DQ, OPT, LINEAR, NN and GAUSS. In order to assess the LFI quality, the pair-wise comparison (PC) method with a two-alternative-forced-choice is carried out and the just-objectionable-differences value is provided, which is similar to the difference-mean-opinion-score value.

Fig. \ref{figdatabase}(e) and Fig. \ref{figdatabase}(f) shows the original images and their distribution of SI and CF for the SMART database. This database is composed of 16 original LFIs and 256 distorted sequences are obtained by introducing 4 compression distortions which include HEVC Intra, JPEG, JPEG2000 as well as Sparse Set and Disparity Coding. Similarly, the PC method is exploited to collect the subjective ratings and the Bradley-Terry scores are provided.

The VALID database has 5 reference LFIs and 40 distorted LFIs under 5 compression artifacts. Fig. \ref{figdatabase}(g) and Fig. \ref{figdatabase}(h) shows the original images and the corresponding SI and CF distribution. Note that the VALID database includes both 8bit and 10bit LFIs. The comparison-based adjectival categorical judgement methodology is used to 8bit images, while the double stimulus impairment scale is performed for 10bit images. In addition, the MOS values are provided for the LFIs.

\renewcommand\arraystretch{1.3}
\begin{table}[!htb]
\centering
\caption{Performance Results of Synthetic and Real Scenes for Tensor-NLFQ on Win5-LID Database.}
\begin{tabular}{c|ccc}
\hline
\textbf{Methods} & \textbf{SRCC} & \textbf{LCC} & \textbf{RMSE} \\ \hline
\textbf{Synthetic scene} &0.8837 &0.9314 &0.3339 \\
\textbf{Real scene}	     &0.8834 &0.9144 &0.3429 \\ \hline
\end{tabular}
\label{data}
\end{table}

\renewcommand\arraystretch{1.3}
\begin{table}[!htb]
	\centering
	\scriptsize
	\caption{Comparison Results of Tucker Decomposition and PCA on Win5-LID Databse.}
	\begin{tabular}{c|ccc}
		\hline
		\textbf{Methods} & \textbf{SRCC} & \textbf{LCC} & \textbf{RMSE} \\ \hline
        PCA                      &0.8858 &0.8962 &0.4166 \\
        \textbf{Tucker decomposition (Proposed)} &\textbf{0.9101} &\textbf{0.9217} &\textbf{0.3781} \\ \hline
	\end{tabular}
    \label{tucker}
\end{table}

To evaluate the model performance on these databases, we choose four evaluation criteria, including Spearman rank-order correlation coefficient (SRCC), linear correlation coefficient (LCC), root mean square error (RMSE) and outlier ratio (OR). The SRCC measures the monotonicity, while LCC focuses on the linear relationship. The RMSE and OR provide the measure of prediction accuracy and consistency, respectively. Higher SRCC and LCC values as well as lower RMSE and OR values represent better performance. Before computing LCC, RMSE and OR, a nonlinear function is adopted as:

\begin{equation}
f(q)=\beta_1\{\frac{1}{2}-\frac{1}{1+exp[\beta_2(q-\beta_3)]}\}+\beta_4q+\beta_5
\end{equation}
where $q$ is the output of a specific objective metric. The parameters $\beta_{1 \cdots 5}$  are optimized to minimize the given goodness-of-fit measure.

\renewcommand\arraystretch{1.3}
\begin{table}[!htb]
	\centering
	\caption{Performance of Individual Color Channels on Win5-LID and MPI-LFA Databases.}
	\begin{tabular}{p{1.2cm}<{\centering}|p{0.8cm}<{\centering} p{0.8cm}<{\centering} p{0.7cm}<{\centering}|p{0.8cm} p{0.8cm} p{0.6cm}}
		\hline
		& \multicolumn{3}{c|}{\textbf{Win5-LID}}       & \multicolumn{3}{c}{\textbf{MPI-LFA}}   \\ \hline
		\textbf{Channel} & \textbf{SRCC}  & \textbf{LCC}    & \textbf{RMSE}  & \textbf{SRCC}  & \textbf{LCC}    & \textbf{RMSE} \\ \hline
		\textbf{$L$} & 0.8693          & 0.8838       & 0.4438       &0.8997     &0.9066     &0.8341 \\
		\textbf{$a^*$}  & 0.7574         & 0.7638        & 0.6210       &0.8299 &0.8468 &1.0367          \\
		\textbf{$b^*$} & 0.8484         & 0.8592        & 0.5038       &0.8883 &0.8845 &0.9197          \\ \hline
\textbf{Proposed} & \textbf{0.9101} & \textbf{0.9217} & \textbf{0.3781} & \textbf{0.9221} & \textbf{0.9294} & \textbf{0.7241} \\ \hline
	\end{tabular}
	\label{tbcolor}
\end{table}

\renewcommand\arraystretch{1.3}
\begin{table}[!htb]
	\centering
	\caption{Performance of Four Direction View Stacks on Win5-LID Database.}
	\begin{tabular}{c|ccc}
		\hline
		& \multicolumn{3}{c}{\textbf{Win5-LID}}          \\ \hline
		\textbf{Orientation} & \textbf{SRCC}  & \textbf{LCC}    & \textbf{RMSE}   \\ \hline
		\textbf{Horizontal} & 0.8850          & 0.8994       & 0.3947        \\
		\textbf{Vertical}  & 0.8529         & 0.8653        & 0.4541               \\
		\textbf{Left diagnoal} & 0.8795         & 0.8942        & 0.3944               \\
		\textbf{Right diagnoal} & 0.8819         & 0.8894        & 0.4271               \\  \hline
\textbf{Proposed} & \textbf{0.9101} & \textbf{0.9217} & \textbf{0.3781} \\ \hline
	\end{tabular}
	\label{tbdirection}
\end{table}

\renewcommand\arraystretch{1.3}
\begin{table}[!htb]
	\centering
	\caption{Performance of Different Structure Similarity Methods on Win5-LID and MPI-LFA Databases.}
	\begin{tabular}{p{1.9cm}<{\centering}|p{0.7cm}<{\centering} p{0.7cm}<{\centering} p{0.7cm}<{\centering}|p{0.6cm}<{\centering} p{0.6cm}<{\centering} p{0.6cm}<{\centering}}
		\hline
		& \multicolumn{3}{c|}{\textbf{Win5-LID}}       & \multicolumn{3}{c}{\textbf{MPI-LFA}}   \\ \hline
		\textbf{Method} & \textbf{SRCC}  & \textbf{LCC}    & \textbf{RMSE}  & \textbf{SRCC}  & \textbf{LCC}    & \textbf{RMSE} \\ \hline
		\textbf{SSIM \cite{wang2004image}} & 0.9101          & 0.9217       & 0.3781       &0.9221     &0.9294     &0.7241 \\
		\textbf{MS-SSIM \cite{wang2003multiscale}}  & 0.9026         & 0.9159        & 0.3755       &0.9305 &0.9300 &0.7282          \\
		\textbf{FSIM \cite{zhang2011fsim}} & 0.8876         & 0.9087        & 0.4029       &0.9227 &0.9268 &0.7743          \\
		\textbf{IW-SSIM \cite{wang2011information}} & 0.8981         & 0.9149        & 0.3834       &0.9268 &0.9288 &0.7368          \\ \hline
	\end{tabular}
	\label{tbss}
\end{table}

Additionally, each database is randomly divided into 80\% for training and the remaining 20\% for testing. We perform 1000 iterations of cross validation on each database. We also provide the median SRCC, LCC, RMSE and OR values as the final measurement.

\subsection{Comparison with Other Objective Metrics}
In order to prove the effectiveness of our proposed Tensor-NLFQ model, we conduct fully experiments by using existing 2D, 3D image, multi-view and LFI quality assessment algorithms. Specifically, we compare with ten 2D FR IQA metrics \cite{wang2004image,sheikh2006image,zhang2011fsim,wang2003multiscale,wang2011information,chandler2007vsnr,sheikh2005information,damera2000image,mantiuk2011hdr}, three 2D NR IQA metrics \cite{mittal2012making,mittal2012no,ghadiyaram2017perceptual}, one 3D FR IQA metric \cite{chen2013full}, two 3D NR IQA metrics \cite{chen2018blind,liu2017binocular}, five multi-view FR IQA metrics \cite{battisti2015objective,sandic2015dibr1,sandic2015dibr,sandic2016multi}, one multi-view NR IQA metric \cite{gu2018model}, one RR LFI quality assessment metric \cite{8632960}, and one NR LFI quality assessment metric \cite{shi2019belif}.

TABLE \ref{table_overall} shows the overall performance of state-of-the-art objective models on the Win5-LID, MPI-LFA and SMART databases, where bold values indicate the best performance results. In TABLE \ref{table_overall}, the FR approaches are modeled by formulas, while NR methods are supervised learning algorithms except for NIQE \cite{mittal2012making}. Furthermore, all these learning-based methods are trained by the same percentages with the proposed algorithm on each database. As shown in this table, our proposed Tensor-NLFQ achieves superior performance compared with state-of-the-art algorithms. One possible explanation is that existing 2D and 3D IQA approaches only focus on spatial quality rather than angular consistency. Although multi-view IQA metrics consider distortion caused by angular interpolation, they aim to deal with the hole distortion caused by the synthesis. Thus, it is not possible to effectively measure LFI artifacts, such as compression distortion. The LFI-IQM \cite{8632960} method ignores the spatial texture information. Moreover, the LFI-IQM \cite{8632960} is influenced by depth map estimation, while BELIF \cite{shi2019belif} cannot take into account the chrominance effects and diverse directions of LFIs. Therefore, their performance is worse than that of the proposed method.

\renewcommand\arraystretch{1.3}
\begin{table}[!htb]
	\centering
	\caption{Performance of Proposed Quality Components on Win5-LID and MPI-LFA Databases.}
	\begin{tabular}{p{1.2cm}<{\centering}|p{0.8cm}<{\centering} p{0.8cm}<{\centering} p{0.7cm}<{\centering}|p{0.8cm} p{0.8cm} p{0.6cm}}
		\hline
		& \multicolumn{3}{c|}{\textbf{Win5-LID}}       & \multicolumn{3}{c}{\textbf{MPI-LFA}}   \\ \hline
		\textbf{Features} & \textbf{SRCC}  & \textbf{LCC}    & \textbf{RMSE}  & \textbf{SRCC}  & \textbf{LCC}    & \textbf{RMSE} \\ \hline
		\textbf{$\textbf{f}_{PCSC}$} & 0.8001          & 0.8188       & 0.5821       &0.8749     &0.8815     &0.8345 \\
		\textbf{$\textbf{f}_{TAVI}$} & 0.8318         & 0.8521        & 0.4648       &0.7912     &0.7964     &1.1669 \\ \hline
\textbf{Proposed} & \textbf{0.9101} & \textbf{0.9217} & \textbf{0.3781} & \textbf{0.9221} & \textbf{0.9294} & \textbf{0.7241} \\ \hline
	\end{tabular}
	\label{tbfeats}
\end{table}

\renewcommand\arraystretch{1.3}
\begin{table}[!htb]
	\centering
	\caption{Cross Validation Results. We Train Our Proposed Model on Win5-LID and Test on MPI-LFA.}
	\begin{tabular}{c|ccc}
		\hline
		\textbf{}    & \textbf{SRCC} & \textbf{LCC}  & \textbf{RMSE} \\ \hline
		\textbf{Proposed} & 0.8469         & 0.8192              & 0.3282        \\ \hline
	\end{tabular}
\label{tbcross}
\end{table}

Further, we provide the performance comparison on the VALID database which includes both 8bit and 10bit LFIs. The VALID database only has 5 original LFIs, whose SI and CF distribution is relatively concentrated. As we can see from TABLE \ref{table_valid}, the proposed Tensor-NLFQ delivers good performance values and especially outperforms state-of-the-art NR IQA algorithms for 10bit LFIs.

To illustrate the prediction results more clearly, the scatter plots of two existing metrics and the proposed model on the Win5-LID and MPI-LFA databases are shown in Fig. \ref{figscatter}. Since the points of our proposed method are more centralized than that of the other metrics, the predictions of our proposed Tensor-NLFQ are more consistent with subjective quality scores. In addition, Fig. \ref{figpercentages} shows the change of performance results for our proposed Tensor-NLFQ method with respect to the training and testing percentages. We can observe that a large number of training data generally bring about the increase of SRCC and LCC performance on both Win5-LID and MPI-LFA databases, which is consistent with \cite{chen2018blind}. Moreover, we choose 80\%-20\% for the training-testing split since this is a common practice in quality assessment \cite{moorthy2011blind,saad2012blind}.

Besides direct performance comparisons, we also quantitatively evaluate the statistical significance using the t-test \cite{mittal2012no} based on the SRCC values obtained from 1000 train-test trials. Here, the null hypothesis is that the mean correlation for the proposed method is equal to that for the compared state-of-the-art algorithm with a confidence of 95\%. The experimental results demonstrate that our proposed Tensor-NLFQ significantly outperforms state-of-the-art objective IQA algorithms.

\subsection{Robustness Against Distortion and Data Types}
\renewcommand\arraystretch{1.3}
\begin{table*}[!htb]
	
	\centering
	\scriptsize
	\caption{Performance Comparison of The Computation Time against SRCC, LCC, RMSE and OR on Win5-LID Database.}
	\begin{tabular}{c|c|c|cccc}
		
		\hline
\textbf{Type}	 & \textbf{Metrics}     & \textbf{Total Computation Time (s)}   & \textbf{SRCC}  & \textbf{LCC}   & \textbf{RMSE} & \textbf{OR} \\ \hline
		
\multirow{10}{*}{\textbf{2D FR}}	 & \textbf{PSNR}     	&0.8188     & 0.6026          & 0.6189          & 0.8031    &0.0045        \\
		
	 & \textbf{SSIM \cite{wang2004image}}      							&2.3068    & 0.7346          & 0.7596          & 0.6650     &0.0000    \\
		
	 & \textbf{MS-SSIM \cite{wang2003multiscale}}    				&3.2937     & 0.8266          & 0.8388          & 0.5566    &0.0000     \\
		
	& \textbf{FSIM \cite{zhang2011fsim}}     							   &14.4056     & 0.8233          & 0.8318          & 0.5675    &0.0045     \\
		
	& \textbf{IW-SSIM \cite{wang2011information}}    				 &27.0113    & 0.8352          & 0.8435          & 0.5492   &0.0000    \\
		
	& \textbf{IFC \cite{sheikh2005information}}     					&69.5933      & 0.5028          & 0.5393          & 0.8611  &0.0000   \\
		
	 & \textbf{VIF \cite{sheikh2006image}}      							   &65.9176     & 0.6665          & 0.7032          & 0.7270   &0.0000  \\
		
	& \textbf{NQM \cite{damera2000image}}     						   &16.5390      & 0.6508          & 0.6940          & 0.7362  &0.0045    \\
		
	 & \textbf{VSNR \cite{chandler2007vsnr}}     						 &4.0910    & 0.3961          & 0.5050          & 0.8826  &0.0182  \\
		
	& \textbf{HDR-VDP2 \cite{mantiuk2011hdr}}  						  &115.1300 		&0.5555 &0.6300 &0.7941 &0.0045  \\ \hline
		
\multirow{3}{*}{\textbf{2D NR}}	& \textbf{BRISQUE \cite{mittal2012no}}    &4.4593   & 0.6687          & 0.7510          & 0.5619  &0.0000   \\
		
	& \textbf{NIQE \cite{mittal2012making}}     &8.8498     & 0.2086          & 0.2645          & 0.9861   &0.0045   \\
		
	& \textbf{FRIQUEE \cite{ghadiyaram2017perceptual}}      &2343.0336   & 0.6328          & 0.7213          & 0.5767  &0.0000      \\ \hline
		
\textbf{3D FR}	& \textbf{Chen \cite{chen2013full}}  &1239.3772        & 0.5269          & 0.6070          & 0.8126  &0.0091    \\ \hline
		
\multirow{2}{*}{\textbf{3D NR}}	& \textbf{SINQ \cite{liu2017binocular}}     &309.7299     & 0.8029          & 0.8362          & 0.5124  &0.0000    \\
		
	& \textbf{BSVQE \cite{chen2018blind}}    &396.6745     & 0.8179          & 0.8425          & 0.4801  &0.0000       \\ \hline
		
\multirow{5}{*}{\textbf{Multi-view FR}}	& \textbf{MP-PSNR Full \cite{sandic2015dibr}}  &32.2917		& 0.5335          & 0.4766          & 0.8989   &0.0000       \\
		
	& \textbf{MP-PSNR Reduc \cite{sandic2016multi}} &16.2708	& 0.5374          & 0.4765          & 0.8989   &0.0000      \\
		
	& \textbf{MW-PSNR Full \cite{sandic2015dibr1}} &1.1421	 & 0.5147          & 0.4758          & 0.8993   &0.0000       \\
		
	& \textbf{MW-PSNR Reduc \cite{sandic2015dibr1}} &1.1352		& 0.5326          & 0.4766          & 0.8989   &0.0000     \\
		
	& \textbf{3DSwIM \cite{battisti2015objective}}   &322.9451     & 0.4320          & 0.5262          & 0.8695   &0.0182     \\ \hline
		
\textbf{Multi-view NR}	& \textbf{APT \cite{gu2018model}}     &2626.8449      & 0.3058          & 0.4087          & 0.9332  &0.0045   \\ \hline
		
\textbf{LFI RR}	& \textbf{LF-IQM \cite{8632960}}     &1168.7424      & 0.4503          & 0.4763          & 0.8991  &0.0273   \\ \hline
		
\textbf{LFI NR}	& \textbf{Proposed NR-LFQA}     & 865.0019     & \textbf{0.9032} & \textbf{0.9206} & \textbf{0.3876} & \textbf{0.0000} \\ \hline
	\end{tabular}
\label{complexity}
\end{table*}

Since the Win5-LID and MPI-LFA involve various distortion types, it is interesting to know how our proposed model performs for individual distortion types. The performance results for each separate distortion type are listed in TABLE \ref{tbsingledis}. Due to the space constraints, we only show RMSE results. It can be seen that our proposed Tensor-NLFQ method outperforms existing objective metrics for most distortion types. Moreover, the proposed model achieves the best performance for typical reconstruction distortions because the reconstruction distortion mainly destroys angle consistency and usually has little influence on spatial quality. Therefore, existing IQA models are difficult to handle such distortions.

Although HEVC compression distortion and Gaussian blur mainly cause the degradation of spatial quality, our Tensor-NLFQ is still very competitive and has a good performance. The JPEG distortion in Win5-LID is introduced based on lenslet, and it affects both spatial quality and angular consistency of LFI. The proposed method considers the effects of both two factors, it is thus not surprising that our model obtains the best performance for JPEG distorted LFIs. Overall, the proposed Tensor-NLFQ can achieve promising performance against existing objective evaluation algorithms regarding to various distortion types.

In addition to different distortion types, we test the performance of our proposed Tensor-NLFQ method on synthetic and real scenes separately. As shown in TABLE \ref{data}, the proposed Tensor-NLFQ can handle both synthetic and real scenes.

\subsection{Validity of Tucker Decomposition}
According to \cite{kolda2009tensor,lee2014incremental}, the Tucker decomposition can be regarded as the higher-order generalizations of PCA or SVD. Since the SVD operation generates huge matrices that make computation difficult, we thus adopt the PCA for performance comparison. Specifically, we reshape LFIs as matrices and then use PCA. The experimental results are shown in TABLE \ref{tucker}. From TABLE \ref{tucker}, we can see that our proposed Tucker decomposition outperforms PCA. This is because the tensor-based approach preserves the correlation of spatial information.

\subsection{Validity of Individual Color Channel}
Since the luminance and chrominance features of LFI are utilized in the proposed model, it is necessary to know how much contribution each color channel has. TABLE \ref{tbcolor} exhibits the performance of individual color channels on Win5-LID and MPI-LFA databases. It can be seen that the luminance channel achieves the best performance among three color channels, which proves that the luminance has the most important influence on LFI quality. Moreover, it is observed that two chrominance channels deliver good performance on two databases, which demonstrates that chrominance also has a significant impact on LFI quality.

\subsection{Validity of Single Orientation View Stack}
In the proposed Tensor-NLFQ method, we weight the extracted features of four orientation view stacks to predict the LFI overall quality. It is meaningful to verify the performance of the view stack in each orientation. Since the MPI-LFA database only includes the horizontal view stack, we present the results on the Win5-LID database, as shown in TABLE \ref{tbdirection}. We can observe that for 4D LFIs, the view stack in each direction has a good performance, which indicates that the characteristics of each orientation can reflect the LFI quality to some extent. Meanwhile, the performance of the final model with four directional feature weighting is significantly improved.

\renewcommand\arraystretch{1.3}
\begin{table}[!htb]
\centering
\caption{Performance Results of The Computation Time for Tensor-NLFQ with Different LFI dimensions on Win5-LID Database.}
\begin{tabular}{c|c}
\hline
\textbf{$\mathbf{L}(s,t,x,y)$} & \textbf{Total Computation Time (s)} \\ \hline
\textbf{$\mathbf{L}(9,9,434,625)$}	&865.0019 \\
\textbf{$\mathbf{L}(5,5,434,625)$}	&361.7581 \\
\textbf{$\mathbf{L}(3,3,434,625)$}	&186.2238 \\
\textbf{$\mathbf{L}(9,9,217,313)$}	&200.4934 \\
\textbf{$\mathbf{L}(9,9,145,208)$}	&99.4762 \\
\textbf{$\mathbf{L}(5,5,217,313)$}	&94.0500 \\ \hline
\end{tabular}
\label{resolution}
\end{table}

\subsection{Different Structure Similarity Methods}
In the TAVI measurement, we use SSIM \cite{wang2004image} as an algorithm for measuring structure similarity. Except for SSIM, several variants of SSIM have been proposed, such as MS-SSIM \cite{wang2003multiscale}, FSIM \cite{zhang2011fsim} and IW-SSIM \cite{wang2011information}. Therefore, we wonder how the proposed method performs when we adopt these algorithms. TABLE \ref{tbss} illustrates the results of our proposed model using different structure similarity methods on Win5-LID and MPI-LFA databases, which indicates that our Tensor-NLFQ model does not rely much on specific structural similarity algorithms.

\subsection{Validity of Individual Proposed Feature}
In this section, we explore the validity of two proposed features (i.e. $\textbf{f}_{PCSC}$ and $\textbf{f}_{TAVI}$) of our model. The performance values of these two features are shown in TABLE \ref{tbfeats}. It can be seen that $\textbf{f}_{PCSC}$ has a good performance on both databases due to the effectiveness of measuring spatial quality, especially on the MPI-LFA database. The reason may be that some interpolation operations in the MPI-LFA database can also cause the degradation of spatial quality. Therefore, in addition to capture the deterioration of spatial quality, $\textbf{f}_{PCSC}$ can also measure a certain degree of angular distortion. Furthermore, $\textbf{f}_{TAVI}$ delivers the slightly lower performance on the MPI-LFA database. Since the MPI-LFA database contains many angular distorted LFIs with low distortion levels and the quality difference is small, human is insensitive to these sequences. However, the proposed $\textbf{f}_{TAVI}$ can capture the degradation of angular consistency effectively. This phenomenon can be shown in Fig. \ref{figSS}, where the curve of distortion sequences are significantly different. Overall, the results validate our proposed features and the performance is improved after the feature combination.

\subsection{Model Generality and Time Complexity}
To validate the model generality, we choose the same distortion in the MPI-LFA and Win5-LID databases to conduct experiments. Specifically, we train the proposed Tensor-NLFQ on the Win5-LID database, and then test it on the LFIs with the same distortion in the MPI-LFA database. The results are shown in TABLE \ref{tbcross}. We can observe that the proposed model is independent for the adopted database.

In addition, we compare the proposed Tensor-NLFQ method with state-of-the-art quality assessment approaches for computational complexity on the Win5-LID database. It should be noted that we also list the SRCC, LCC, RMSE and OR for fair comparison. As shown in TABLE \ref{complexity}, our proposed Tensor-NLFQ is demonstrated to have lower computation time compared to LF-IQM \cite{8632960}. The reason may be that different from conventional LFI quality assessment metrics, our proposed approach relieves the complex computation for estimating depth maps. In general, our proposed Tensor-NLFQ method is in the same level of time complexity compared with state-of-the-art 3D quality assessment metrics and demonstrates the best SRCC, LCC, RMSE and OR performance among all algorithms. Further, we analyze the effects of different angular and spatial dimensions on time complexity. Specifically, we sample the LFI in angular and spatial dimensions by $1/2$ and $1/3$ times separately. The LFI is also sampled in angular and spatial dimensions by $1/2$ times simultaneously. The performance results are shown in TABLE \ref{resolution}. We can see that the time complexity of our proposed Tensor-NLFQ is affected by the angular and spatial dimensions of LFIs. The reduction of angular and spatial dimensions significantly reduces the computation time.

\section{Conclusion}
In this paper, we present a novel Tensor oriented No-reference Light Field image Quality evaluator (Tensor-NLFQ). According to the existing research and our previous work, color information has a significant impact on the perceived LFI quality. We thus introduce luminance and chrominance information in our proposed model. Since the LFI can be regarded as a high-dimensional tensor signal, we exploit the tensor decomposition to extract the principal components of LFI, which can effectively reflect the LFI quality. The angular consistency of diverse directions is considered in the proposed method, including horizontal, diagonal, vertical and right diagonal orientations. As the LFI quality is affected by both spatial quality and angular consistency, we propose principal component spatial characteristic and tensor angular variation index to measure the degradation of spatial quality and angular consistency, respectively. We conduct extensive experiments to compare the proposed Tensor-NLFQ with existing 2D, 3D image, multi-view and LFI quality assessment algorithms. The results demonstrate that our approach outperforms state-of-the-art metrics and can handle the typical distortions of LFI.

In the future, we will extend the proposed model to light field video quality assessment. Moreover, how to apply our proposed method to the optimization of existing image compression and reconstruction algorithms could also be further explored.


%





\ifCLASSOPTIONcaptionsoff
  \newpage
\fi



%
\bibliographystyle{IEEEtran}
\bibliography{tipRefer}

\end{document}